\documentclass{ieeeaccess}
\usepackage{cite}
\usepackage{amsmath,amssymb,amsfonts}
\usepackage{algorithmic}
\usepackage{graphicx}
\usepackage{textcomp}
\usepackage{array}
\usepackage{rotfloat}
\usepackage[font={sf,scriptsize},           
            labelfont={bf,color=accessblue},
            caption=false]{subfig} 
\usepackage[dvipsnames]{xcolor}

\newif\ifheavyimages
\heavyimagestrue 

\newcolumntype{P}[1]{>{\centering\arraybackslash}p{#1}}
\newcolumntype{M}[1]{>{\centering\arraybackslash}m{#1}}
 
\def\BibTeX{{\rm B\kern-.05em{\sc i\kern-.025em b}\kern-.08em
    T\kern-.1667em\lower.7ex\hbox{E}\kern-.125emX}}
    
\makeatother

\begin{document}

\title{Dash Sylvereye: A WebGL-powered Library for Dashboard-driven Visualization of Large Street Networks}

\author{\uppercase{Alberto Garcia-Robledo}\authorrefmark{1}, 
\uppercase{Mahboobeh Zangiabady\authorrefmark{2}}}
\address[1]{Conahcyt, CentroGeo, Quer\'etaro, M\'exico (e-mail: agarcia@centrogeo.edu.mx)}
\address[2]{University of Twente, Enschede, the Netherlands (e-mail: m.zangiabady@utwente.nl)}

\corresp{Corresponding author: Alberto Garcia-Robledo (e-mail: agarcia@centrogeo.edu.mx).}

\begin{abstract}
State-of-the-art open network visualization tools like Gephi, KeyLines, and Cytoscape are not suitable for studying street networks with thousands of roads since they do not support simultaneously polylines for edges, navigable maps, GPU-accelerated rendering, interactivity, and the means for visualizing multivariate data. To fill this gap, the present paper presents Dash Sylvereye: a new Python library to produce interactive visualizations of primal street networks on top of tiled web maps. Thanks to its integration with the Dash framework, Dash Sylvereye can be used to develop web dashboards around temporal and multivariate street data by coordinating the various elements of a Dash Sylvereye visualization with other plotting and UI components provided by the Dash framework. Additionally, Dash Sylvereye provides convenient functions to easily import OpenStreetMap street topologies obtained with the OSMnx library. Moreover, Dash Sylvereye uses WebGL for GPU-accelerated rendering when redrawing the road network. We conduct experiments to assess the performance of Dash Sylvereye on a commodity computer when exploiting software acceleration in terms of frames per second, CPU time, and frame duration. We show that Dash Sylvereye can offer fast panning speeds, close to 60 FPS, and CPU times below 20 ms, for street networks with thousands of edges, and above 24 FPS, and CPU times below 40 ms, for networks with dozens of thousands of edges. Additionally, we conduct a performance comparison against two state-of-the-art street visualization tools. We found Dash Sylvereye to be competitive when compared to the state-of-the-art visualization libraries Kepler.gl and city-roads. Finally, we describe a web dashboard application that exploits Dash Sylvereye for the analysis of a SUMO vehicle traffic simulation.
\end{abstract}


\titlepgskip=-15pt

\maketitle


\label{sec:introduction}

\PARstart{O}{ne} of the primary objects of interest of urban researchers and planners are street networks. They study street networks for a variety of applications such as traffic engineering, transportation, and urban planning. Recently, academics from seemingly unrelated fields, such as Networks Science and Computer Science, have joined to study the complexity of large street networks and develop efficient algorithms to process them. 

With the advent of the OpenStreetMap (OSM) project \cite{Haklay2008} the street topology of virtually any city in the world became publicly available for analysis. Tools like OSMnx \cite{Boeing2017, Boeing2017b} make it easy for any researcher to download OSM street network data with a simple query. However, the availability of such networks has also revealed the limitations of current tools like graph visualization. 

Urban researchers need tools to make sense of multivariate data associated with street networks. These data are hardly static: vehicle counts, vehicle positions, traffic bottlenecks, and other urban data change over time. Dashboards have become a standard visual analytics tool when trying to make sense of multivariate data. Prominent dashboard tools in the industry include Tableau \cite{Batt2020} and Google Data Studio\footnote{https://datastudio.google.com/}. However, these kinds of open tools are too general to support the practical analytical needs of real-world urban applications \cite{Zheng2016}. 

On the other hand, the street network of large cities is made of dozens of thousands of nodes and edges. This imposes the need to push the processing capabilities of graphics adapters to render such large structures. These complex visualizations should also allow for user interactivity by enabling navigation, panning, zooming, and clicking. 

The use of web technologies for developing visualization solutions is currently a tendency among practitioners. Open-source programming libraries like the Dash framework enable data analysts to develop their own rich and interactive web dashboards by exploiting a variety of coordinated web plotting and UI components. Such dashboards can be displayed in any modern web browser and be easily deployed on the web. 

State-of-the-art graph visualization tools like Gephi, KeyLines, and Cytoscape are not suitable for studying city-scale street networks since they do not support simultaneously polyline\footnote{A sequence of connected segments that describe a curve.} drawing for edges, navigable maps, interactivity, the means for visualizing multivariate data, and GPU-accelerated rendering. 

In this context, the research question we tackle is: how to fill this gap and exploit state-of-the-art visual analytics techniques and web technologies to produce interactive visualizations of street networks on a city scale, along with its multivariate data, making this technology easily available to researchers and practitioners alike? 

To answer the posed question, this paper presents a new Python library called Dash Sylvereye which produces interactive visualizations of primal street networks on top of tiled web maps. Thanks to its integration with the Dash framework, Dash Sylvereye can be easily exploited to develop web dashboards around temporal and multivariate urban data by coordinating the various elements of a Dash Sylvereye visualization with other Dash plotting and UI components.  

Dash Sylvereye can render large city-scale interactive street networks as well as thousands of interactive markers in commodity computers with the help of the system’s GPU through WebGL.

The core contributions of this paper are as follows:

\begin{itemize}
    \item A library tool for Python that generates street network visualizations that can draw atop web tile maps and that is designed from the ground up to be compatible with the widely used Dash dashboard visualization framework.
    \item A library tool that allows for the customization of colors, sizes, transparency, and visibility of individual street network elements as well as markers. Visual properties can be also automatically scaled based on the values found in the street network's data.
    \item A library tool that provides fast software acceleration and exploits hardware acceleration for redrawing, showing panning speeds of close to 60 FPS, and CPU times below 20 ms, for street networks with thousands of edges.
\end{itemize}

The rest of this paper is structured as follows. Section \ref{section:background} provides additional background on topics that are relevant to the proposed solution. Section \ref{section:related-work} offers a review of the state-of-the-art on street graph visualization. Section \ref{section:dash-sylvereye-design} lists the requirements we identified were needed to meet and presents details on the internal design of the Dash Sylvereye library. Section \ref{section:usage-example} offers the reader a quick grasp of how coding with the Dash Sylvereye library feels. In Section \ref{section:animation-performance-assessment} we assess the animation performance of Dash Sylvereye in terms of frames per second, frame duration, and CPU time. In Section \ref{section:comparison-to-soa} we present a comparison of the animation performance among Dash Sylvereye and other two state-of-the-art road network visualization libraries. In Section \ref{section:dashboard-example} we describe a non-trivial example of a dashboard that uses Dash Sylvereye as its central component. Finally, in Section \ref{section:conclusion} we offer final conclusions and future work.


\section{Background}\label{section:background}

\subsection{Street networks}

This paper is concerned with primal street graphs. A primal street graph is a non-planar directed multi-graph with loops allowed where nodes represent street intersections or junctions, and edges represent street segments \cite{Boeing2017b}. A street network is a kind of spatial network \cite{Anderson2020}: a graph that models natural, sociological, or technological phenomena where the elements of the graph are mapped to the spatial dimension, usually to geographical coordinates. Urban networks have become the focus of many works in recent years. An example of such works is \cite{Huang2015}, which makes use of a new model for analyzing urban network structures, combining them with the information provided by taxi trajectory data.

\subsection{Web-based visualization}

The wide availability of web browsers has turned them into an all-pervasive execution platform. Recently, an increasing number of web‐based visualization applications have been proposed motivated by the new technologies offered by modern browsers \cite{Mwalongo2016}. The HTML5 standard gives programmers an array of options to render graphics: the HTML canvas, SVG graphics, CSS animations, and WebGL. WebGL is a standardized JavaScript API for rendering GPU-accelerated graphics in web browsers. A WebGL application consists of two parts: control code written in JavaScript and shader code written in the GLSL language. WebGL has particularly attracted the interest of the data visualization community since it allows programmers to exploit the GPU processor regardless of the vendor. 

A good example of a work that exploits state-of-the-art web visualization technologies for graph analysis is ContraNA. ContraNA \cite{Fujiwara2020} is a visual analytics framework that exploits machine learning to compare two networks for learning the main specifications of one network with respect to the other. Such comparison is challenging due to the complex structure of large graphs. The authors developed ContraNA as a web application. The back-end is developed in Python whereas the front-end uses a combination of HTML5, JavaScript, D3.js, and WebGL. WebSockets are used for back-end and front-end communication. Examples of works that exploit WebGL to produce visualizations of large graphs include \cite{Saska2020, Fang2017, Han2021, Li2020}.  

JavaScript has become the \textit{lingua franca} for front-end web development. There already exists a mature ecosystem of open-source JavaScript libraries which are being exploited for data visualization. Prominent examples of such libraries are D3.js and Three.js. More recently, WebAssembly has enabled developers to write high-performance code that rivals in speed with native applications written in C and C++. This technology opens new possibilities for efficiently running compute-intensive algorithms in the browser, such as graph layout algorithms.

\subsection{Dashboard visualization}

Dashboards are one of the most common use cases for data visualization \cite{Sarikaya2018}. Interest in developing web dashboards has recently increased in governments, universities, research centers, and health institutions due to the need of sharing real-time information about the state of the COVID-19 pandemic in an open and accessible manner.

In urban studies, dashboards are being used to visualize real-time urban data from a variety of sources to provide an easy-to-understand tool to decision-makers \cite{Gray2016}. Dashboards can be used to visually assess urban performance to support the sustainable development of smart cities \cite{Jing2019}, and for transparent and accountable decision-making  \cite{Matheus2018}. A good example of a work that exploits dashboards for city analytics purposes is \cite{Sakib2018}, which proposes a dashboard-driven visual tool for analyzing traffic accident and casualty trends.

Python is one of the most-used languages among developers who identify themselves as data scientists \cite{Hayes2020}. There is a relatively new ecosystem of frameworks that are attracting the attention of data science practitioners in need of developing web dashboard applications entirely in Python, without the need of learning front-end web languages like HTML, CSS, and JavaScript. One such library is the Plotly Dash framework\footnote{https://plotly.com/dash/}. 

The Dash framework is built around the concept of \textit{Dash component}. A Dash component is a Python class that provides an abstraction for a web UI element: from a single HTML tag to more complex elements such as a slider, a chart, a gauge meter, or a navigation bar. A Dash component has properties that can be set, read, and updated. Under the hood, Dash components are Python wrappers for components written with the widely-used React.js front-end UI framework\footnote{https://reactjs.org/}. This enables programmers to build their Dash components in JavaScript.

Dash applications are composed of two parts. The first part is the \textit{layout} of the dashboard, which describes the application's appearance. It is specified as a tree of Dash components. The second part is the \textit{callbacks}, which defines the interactivity of the application. Dash callbacks are Python functions that are automatically triggered when the properties of Dash components change. 

A callback receives the values of the changed properties as input and returns new values for other properties as output. Every property of a Dash component can be updated through a callback. A special kind of input is the states: input parameters that do not trigger a callback, only store the state of a parameter at the moment the callback is triggered by another input parameter.

\section{Related work}\label{section:related-work}

Many graph visualization tools have been developed over the last few years to generate graph visualizations. In consequence, the landscape of such tools has become extensive. We focus our state-of-the-art review on both open tools\footnote{Tools that are publicly available as either open-source, free, or commercial products.} and academic works that propose practical contributions with any of the two following features: 1) support for the development of visualization dashboards around the reported tool, 2) some sort of geospatial visualization support, such as the ability to render graphs on top of maps, and 3) rendering of large road networks. We discuss in detail tools that provide any kind of dashboard support in Section \ref{section:tools-with-dashboard-support}.

\begin{table*}
\caption{Feature comparison of state-of-the-art network visualization tools. To grant support for a given feature, it should be provided out-of-the-box or via plug-ins/extensions. A N/A in the \textit{Language} column means that the work is not a library but a web or desktop application. Notes: \textbf{(1)} Support of the feature is granted only if the tool can draw polylines for edges. \textbf{(2)} OSMnx can generate static Web visualizations with Folium. \textbf{(3)} By using the TimestampedGeoJson Folium plugin. \textbf{(4)} By using the Cytoscape.js library. \textbf{(5)} By using the Cytoscape.js Dash component. \textbf{(6)} A KeyLines visualization can be embedded in a Kibi dashboard. \textbf{(7)} There is a Tableau extension for embedding Kepler.gl visualizations. \textbf{(8)} By combining the Dash component with, for example, a Dash slider component that implements a timeline. \textbf{(9)} Since the tools are built on top of the Leaflet.js library, they may be able to display non-OSM tilemaps. \textbf{(10)} It supports OpenCL for accelerating compute-intensive tasks such as graph layout calculation. Cytoscape.js is not powered by WebGL. \textbf{(11)} With Gephi Toolkit. \textbf{(12)} With Cytoscape.js. \textbf{(13)} With PyGraphistry and GraphistryJS. \textbf{(14)} GPU acceleration is used for computing graph layouts. \textbf{(15)} It supports non-georeferenced polylines. \textbf{(16)} It supports JavaScript if the standalone React component is used. However, dashboard framework integration (and by extension timeline support) will not be available. \textbf{(17)} With graph-app-kit. \textbf{(18)} Support for GPU acceleration is mentioned neither in the paper nor webpage. Nonetheless, given that UNA is an extension, GPU acceleration support could be provided through the base packages (ArcGIS and Rhinoceros 3D) which support OpenGL. \textbf{(19} Python support is provided through the keplergl module.\label{table:soa}}.
\begin{centering}
\begin{tabular}{|*{12}{M{1.70cm}|}}
\hline
Tool & Language & Polyline drawing (1) & Rendering on top of a map layer & GPU-accelerated rendering & Programmable interactivity & Timeline support & Dashboard framework integration\tabularnewline
\hline
city-roads & JavaScript & Yes & No & WebGL & No & No & No\tabularnewline
Folium & Python & Yes (2) & OSM, Mapbox, Stamen (9) & No & No & Yes (3) & No\tabularnewline
Gephi & Java (11) & No & No & OpenGL & No & Yes & No\tabularnewline
OSMnx & Python & Yes & OSM (2) & No & No & No & No\tabularnewline
Cytoscape & JavaScript (12) & No & No & No (10) & Yes (4) & Yes (8) & Yes (5)\tabularnewline
KeyLines & JavaScript & No & OSM (9) & WebGL & Yes & Yes & Yes (6)\tabularnewline
Kepler.gl & Python, JavaScript & Yes & Mapbox & WebGL & Yes & Yes & Yes (7)\tabularnewline
Sigma.js & JavaScript & No & No & WebGL & Yes & No & No \tabularnewline
VivaGraphJS & JavaScript & No & No & WebGL & Yes & No & No \tabularnewline
Graphistry & Python, JavaScript (13) & No & No & WebGL & No & Yes (17) & Yes (17) \tabularnewline
ReGraph & JavaScript (19) & No & OSM (9) & WebGL & Yes & Yes & No \tabularnewline
Ogma & JavaScript & No & OSM (9) & WebGL & Yes & Yes & No \tabularnewline
G6 & JavaScript & No (15) & No & No (14) & Yes & No & No \tabularnewline
yFiles & JavaScript & No (15) & OSM (9) & WebGL & Yes & Yes & No \tabularnewline
El Grapho & JavaScript & No & No & GLSL & Yes & No & No \tabularnewline
ngraph.pixel & JavaScript & No & No & GLSL & Yes & No & No \tabularnewline
react-force-graph & JavaScript & No & No & WebGL & Yes & No & No \tabularnewline
ccNetViz & JavaScript & No & No & WebGL & Yes & No & No\tabularnewline
Carina & N/A & No & No & WebGL & No & No & No\tabularnewline
NetV.js & JavaScript & No & No & WebGL & Yes & No & No\tabularnewline
Argo Lite & N/A & No & No & WebGL & No & No & No\tabularnewline
Schoedon et al. 2019 & N/A & Yes & Yes & GLSL & No & No & No\tabularnewline
UNA & N/A & Yes & Yes & ? (18) & No & No & No\tabularnewline
\textbf{D. Sylvereye} & \textbf{Python (16)} & \textbf{Yes} & \textbf {OSM (9)} & \textbf{WebGL} & \textbf{Yes} & \textbf{Yes (8)} & \textbf{Yes}\tabularnewline
\hline 
\end{tabular}
\par\end{centering}
\end{table*}

\subsection{Tools in literature}

The following are works in the academic literature that report graph visualization libraries for a variety of programming languages. We focus our review on tools that can render large road networks (through GPU hardware acceleration).


ccNetViz \cite{Saska2020} is an open-source WebGL-based JavaScript library for network visualization. It supports animation features (nodes and links). Node colors, size, and transparency can be manipulated in real-time. Similarly, the animation of edges can be used to display information transmission. Animation features can be specified dynamically.

Carina \cite{Fang2017} is a visualization tool that helps researchers to explore and visualize large graphs with millions of nodes. Carina supports fast graph drawing through WebGL and supports both desktop (Electron) and mobile platforms. An outstanding feature of Carina is it does not save the whole graph in RAM, enabling the tool to handle networks as big as 69 million edges. 

Authors in \cite{Han2021} developed a visualization tool for large graphs called NetV.js. It is a WebGL-based JavaScript library that supports up to 50 thousand nodes and 1 million edges. It exploits the GPU to enhance the drawing performance and create an interface for manipulating graph components. 

Argo Lite \cite{Li2020} is an interactive network visualization tool for web browsers. Users are enabled to modify the characteristics of nodes (size, shape, colors), links (colors), and labels (size and length). It uses WebGL to draw graphs fast. Users can import graph data from CSV, GEXF, and TSV files.

Authors in \cite{Schoedon2019} developed a web-based application to visualize detailed information of transportation networks for mobility analytics by exploiting reachability maps. It is powered by GLSL.

Urban Network Analysis (UNA) \cite{Sevtsuk2012} is a full-fledged toolbox that can be used to visualize spatial networks, as well as computing network measurements. It is provided as an extension for ArcGIS and Rhinoceros 3D. Support for GPU acceleration is not explicitly mentioned in the paper nor webpage. Nonetheless, given that UNA is an extension, GPU acceleration support could be provided through ArcGIS or Rhinoceros 3D.

\subsection{Open tools} 

The following are tools that are made available openly through code repositories across the web. We limited our review on programming libraries that are still active, that show the aforementioned mentioned features, or with an associated programming library. For the sake of comparison with the library reported in this paper, we gave priority to JavaScript and Python libraries, but we also covered the widely-used Gephi tool, which is Java-based. 
 

Folium\footnote{https://github.com/python-visualization/folium} is an open-source Python tool that allows users to visualize data on an interactive Leaflet.js map. Users can zoom or click on the map to analyze the geo-referenced data. 

OSMnx \cite{Boeing2017, Boeing2017b} is an open-source Python library to easily download, visualize, and analyze urban street networks. It is built upon three widely used Python libraries, namely GeoPandas, NetworkX, and Matplotlib. It allows the user to extract street data from OSM for different transport modes such as walking, cycling and driving with a single line of code. OSMnx can also visualize isochrone maps. 

Cytoscape \cite{Shannon2003} is an open-source graph visualization tool originally developed for biological network analysis. Cytoscape provides visualization functions that make it easy for researchers to interactively analyze complex graph datasets. However, it doesn't scale to high-volume graphs. Nonetheless, it supports offloading computationally intensive processing on a GPU, multi-core CPU, or multi-processor card by using OpenCL. 

Gephi \cite{Bastian2009} is an open-source visualization tool for users who seek to generate static visualizations of graphs. It is a desktop application that supports a wide catalog of plug-ins. It is simple to use for beginners. Also, it makes it easy to create CSV files from the network's data. Graphs can be exported to a variety of formats. It is powered by OpenGL. Gephi provides the Gephi Toolkit\footnote{https://gephi.org/toolkit/}, a standalone Java library that programmers can use to generate visualizations programmatically.

Anvaka's city-roads\footnote{https://github.com/anvaka/city-roads} is an open-source visualization web tool written in JavaScript that extracts data from OSM to draw all the streets within a city. It is powered by WebGL.

Sigma.js\footnote{http://sigmajs.org/} is an open-source JavaScript library that supports HTML canvas and WebGL renderers for graph visualization, as well as mouse and touch support. Thanks to its plug-in architecture, the library is extensible. It can import Gephi graphs in GEXF format.

VivaGraphJS\footnote{https://github.com/anvaka/VivaGraphJS} is an open-source JavaScript library that supports WebGL, Canvas, and SVG renderers for graph visualization. It is built on top of the ngraph\footnote{https://github.com/anvaka/ngraph} graph algorithms library. 

ReGraph\footnote{https://cambridge-intelligence.com/regraph/} is a commercial WebGL-powered React library for graph visualization by Cambridge Intelligence. It implements two visualization components: a chart and a time bar.  It can render graphs on top of Leaflet.js web maps. Other geospatial features supported are geo-fencing, overlays, and multiple coordinate reference systems.

Ogma\footnote{https://doc.linkurio.us/ogma/latest/} is a commercial graph visualization JavaScript library by Linkurious. Ogma is powered by WebGL, but it also supports HTML5, Canvas, and SVG renderers. Inserting a custom UI on top of Ogma is possible. Geographical mode allows the programmer to display the graph on top of a web map from different map providers.

G6\footnote{https://g6.antv.vision/en} is a graph visualization JavaScript library. It supports drawing polylines for edges. However, it does not support rendering graphs on top of maps. GPU acceleration is supported for computing graph layouts. 

El Grapho\footnote{https://www.elgrapho.com/} is an open-source JavaScript library for graph visualization that exploits GLSL shaders for quickly generating graph renderings of large graphs. The rendered graphs can be zoomed and panned. It supports multiple graph layout algorithms. Graph renderings in El Grapho are interactive.

ngraph.pixel\footnote{https://github.com/anvaka/ngraph.pixel} is an open-source JavaScript library by the creator of city-roads for visualizing non-road graphs. As city-roads, ngraph.pixel is powered by WebGL. Unlike city-roads, ngraph.pixel allows the programmer to listen to graph change events.

react-force-graph\footnote{https://github.com/vasturiano/react-force-graph} is an open-source WebGL-powered library for graph visualization. react-force-graph is implemented as a React library. Its graph renderer is based on ShaderMaterial from the Three.js 3D JavaScript library\footnote{https://threejs.org/}. It supports both 2D and 3D graph rendering.

\subsection{Tools with dashboard support}\label{section:tools-with-dashboard-support}

The following are graph visualization tools that provide some kind of integration with dashboard visualization frameworks.

Cytoscape.js\footnote{https://js.cytoscape.org/} is a JavaScript library for visualizing and interacting with graphs. It provides a rich set of features and APIs for creating graph visualizations, performing graph analysis, and implementing custom graph algorithms. Cytoscape.js allows the creation and manipulation of nodes and edges, apply various layout algorithms, customizing visual styles, and the handling of user interactions. Cytoscape.js can be integrated into Dash dashboards by exploiting the Dash Cytoscape component\footnote{https://dash.plotly.com/cytoscape}. It does not provide WebGL support.

KeyLines\footnote{http://www.keylines.com} is a commercial JavaScript toolkit for visualizing and interacting with network and graph data. It is powered by WebGL. Since it is neither free nor open-source, users must purchase a license to use it. It supports events to react to user actions such as mouse clicks and drag-and-drop. Kibi, now known as Kibana, is an open-source data exploration and visualization platform primarily built for Elasticsearch. Kibana provides its own set of visualization components and plugins for creating dashboards and exploring data. KeyLines visualizations can be integrated into a Kibi dashboard by utilizing custom development and integration techniques. This may involve embedding KeyLines visualizations within Kibana's dashboard panels or incorporating KeyLines as a separate component within a Kibi dashboard.

Kepler.gl is an open-source geospatial data visualization library. Kepler.gl has the ability to display millions of data points representing numerous trips and perform real-time spatial aggregations by exploiting WebGL. By presenting geospatial data within a unified interface, Kepler.gl enables users to validate concepts and extract insights from these visualizations. Users have the flexibility to visualize spatial datasets with various map layers and explore the data through filtering, animation, and aggregation. The Kepler.gl Tableau extension integrates a Kepler.gl map visualization directly into the Tableau Desktop App, allowing users to interact with the map using the same user interface found in the Kepler.gl demo app. Additionally, the map can be configured to interact with other Tableau charts.

Graphistry is a commercial graph-based analysis tool. It supports WebGL acceleration and provides a Python library called PyGraphistry\footnote{https://github.com/graphistry/pygraphistry} which acts as a client to extract, transform, and load graphs into Graphistry. An alternative Graphistry client is the GraphistryJS JavaScript library\footnote{https://github.com/graphistry/graphistry-js}. Graphistry provides the graph-app-kit \footnote{https://github.com/graphistry/graph-app-kit}, which is a toolkit designed to help build custom graph analytics applications and dashboards. More specifically, the graph-app kit provides integration with a dashboard environment based on the Streamlit library that can be deployed on the cloud. graph-app-kit provides a set of reusable components and utilities to assist with the integration of Graphistry's graph visualization and analysis capabilities into dashboard applications.

\subsection{Discussion}

Table \ref{table:soa} shows a comparison of the discussed network visualization tools and academic works in terms of the following features: 1) programming language (for libraries), 2) support for polyline drawing for edges, 3) support for rendering graphs on top of a map layer, 4) support for GPU-accelerated rendering, 5) support for programmable interactivity, 6) support for a timeline and 7) support for integration with a dashboard framework. We believe that these are the feature a street network visualization tool or library should possess to make it useful for real-world urban street analysis.

As shown in Table \ref{table:soa}, to the best of our knowledge, Dash Sylvereye is the only tool written for Python that generates street network visualizations that can draw atop web tile maps, that supports programmable user interactivity, that exploits hardware acceleration, and, most importantly, that is designed from the ground up to be compatible with a larger dashboard visualization framework. 

When ignoring the target programming language, Kepler.gl is the one tool that holds the most similarities with Dash Sylvereye's feature set. However, unlike Dash Sylvereye, Kepler.gl is not written with dashboard integration as one of its core features.


\section{Dash Sylvereye design}\label{section:dash-sylvereye-design}
 
\subsection{Requirements}

We aim to provide a flexible and accessible tool that allows for the visualization of large road networks with associated multivariate data on commodity systems. This aim involves a series of design requirements:

\begin{itemize}
\item \textit{R1. Support for polyline drawing on top of web tilemaps.} An edge in a road network is defined as a sequence of coordinates that represent its shape in the actual geography. Visualizations should be able to show edges as a sequence of lines given the sequence of coordinates. Also, the road network should be rendered on top of an interactive web tilemap such as those provided by Mapbox or OSM, which allow the user to navigate through the map by panning and zooming.
\item \textit{R2. Support for markers.} Street network visualization is useful for practical applications insofar as it allows for the graphical representation of events that happen around the street network itself, such as traffic warnings, car accidents, and bottleneck spots, as well as places of interest (POI). A common practice in the industry to represent such information in products such as Google Maps and Waze is the use of markers. With this in mind, a street network visualization tool should provide support for drawing customizable markers on top of the map and the road network.
\item \textit{R3. Good frame rate for large street networks on commodity hardware.} The visualization tool should provide an animation frame rate of 24 FPS\footnote{The standard minimum speed needed to experience realistic motion \cite{Bowman2002}.} or higher, for street networks with thousands of nodes and edges, to provide the user a responsive experience when navigating through the visualization (zooming and panning). Such responsive experience should be achievable without the need for a high-end GPU, on commodity hardware such as a laptop computer with a commodity integrated graphics processor (e.g. Intel HD Graphics and AMD Ryzen with Radeon graphics).
\item \textit{R4. Styles for nodes, links, and markers.} The tool should enable the user to customize the visual styles (e.g. color and size) of individual nodes, edges, and markers. Also, it should facilitate the use of the data associated with the street network for styling.
\item \textit{R5. Interactions.} The tool should allow the programmer to listen for events triggered when the user interacts with the elements of the visualization to define custom behavior such as retrieving and showing the data of a clicked node, or showing a popup with custom data on top of a clicked marker.
\item \textit{R6. Support for nodes, edges, and markers to store arbitrary data.} The tool should enable the user to associate arbitrary data with individual elements of the visualization. For example, edges obtained from OSM should be able to store its length, its road type (bridge, highway), maximum speed, etc.
\item \textit{R7. Integration with a dashboard framework.} Most importantly, the tool should enable the street network visualization to work natively with a well-known dashboard framework to allow for the creation of dashboard visualizations of multivariate urban data that complement and enrich the street network visualization.
\end{itemize}

\ifheavyimages
\begin{figure*}
\centering
\subfloat[]{\includegraphics[scale=1.30]{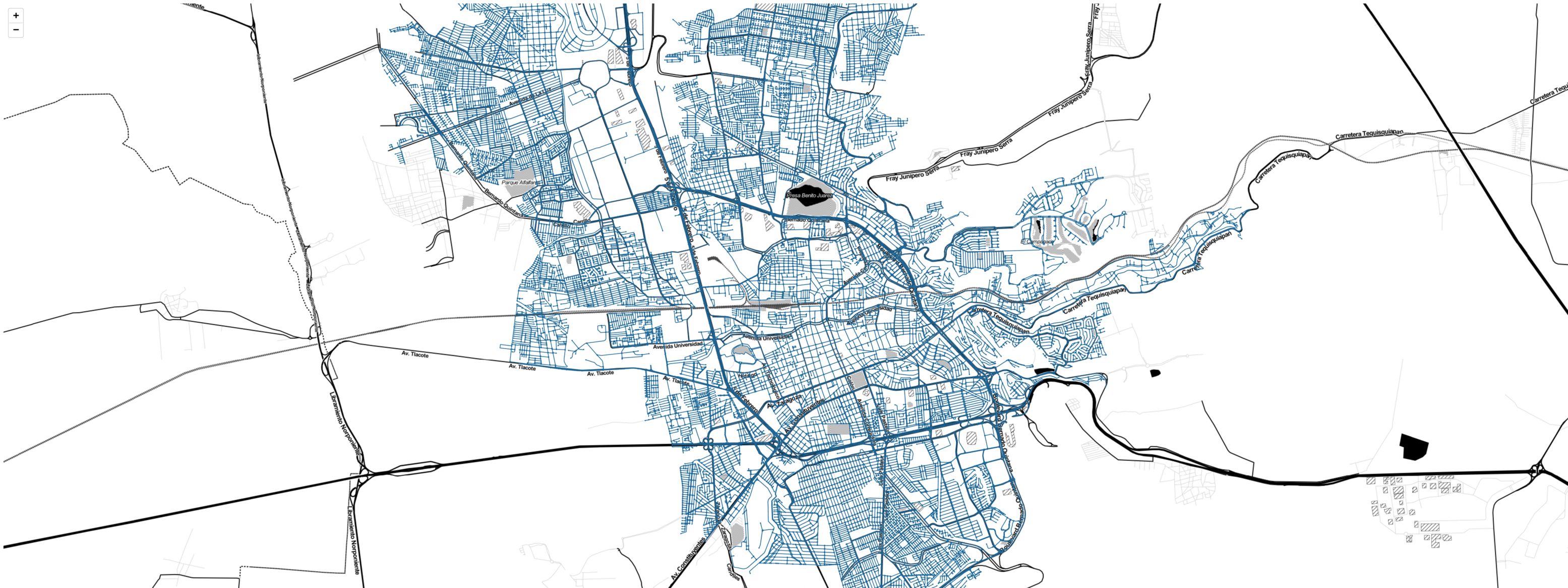}} 

\subfloat[]{\includegraphics[scale=1.30]{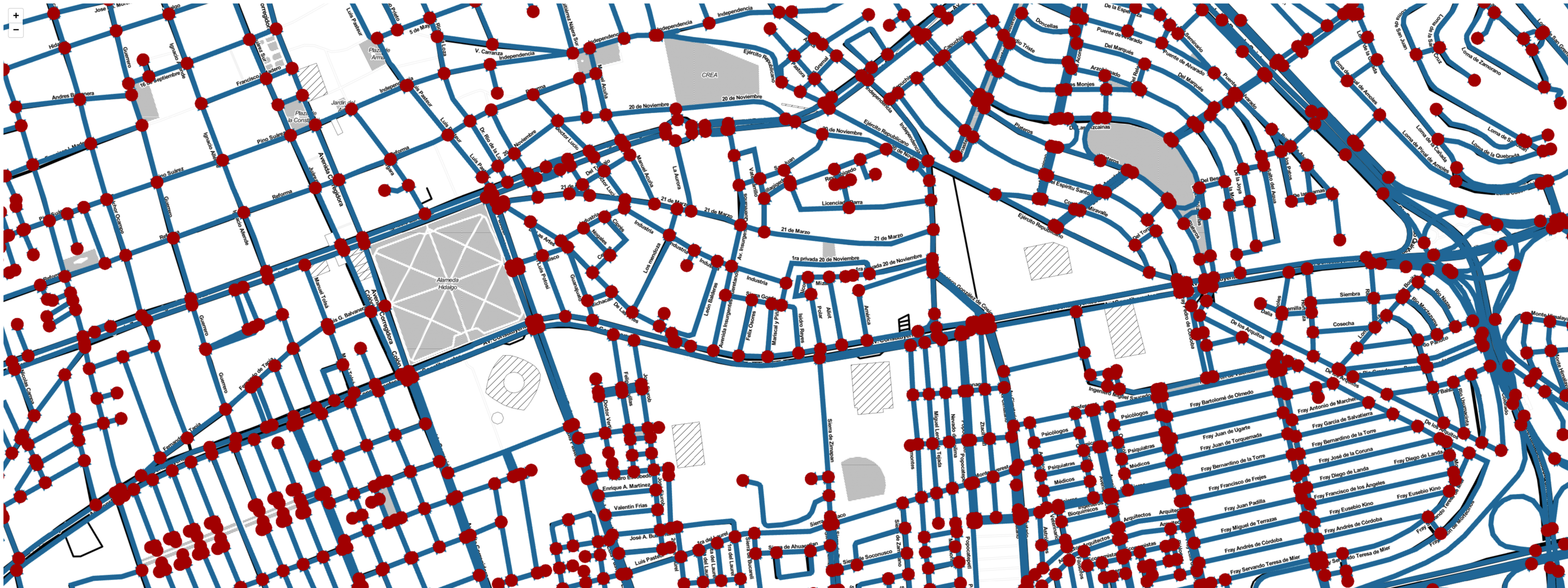}} 

\subfloat[]{\includegraphics[scale=1.30]{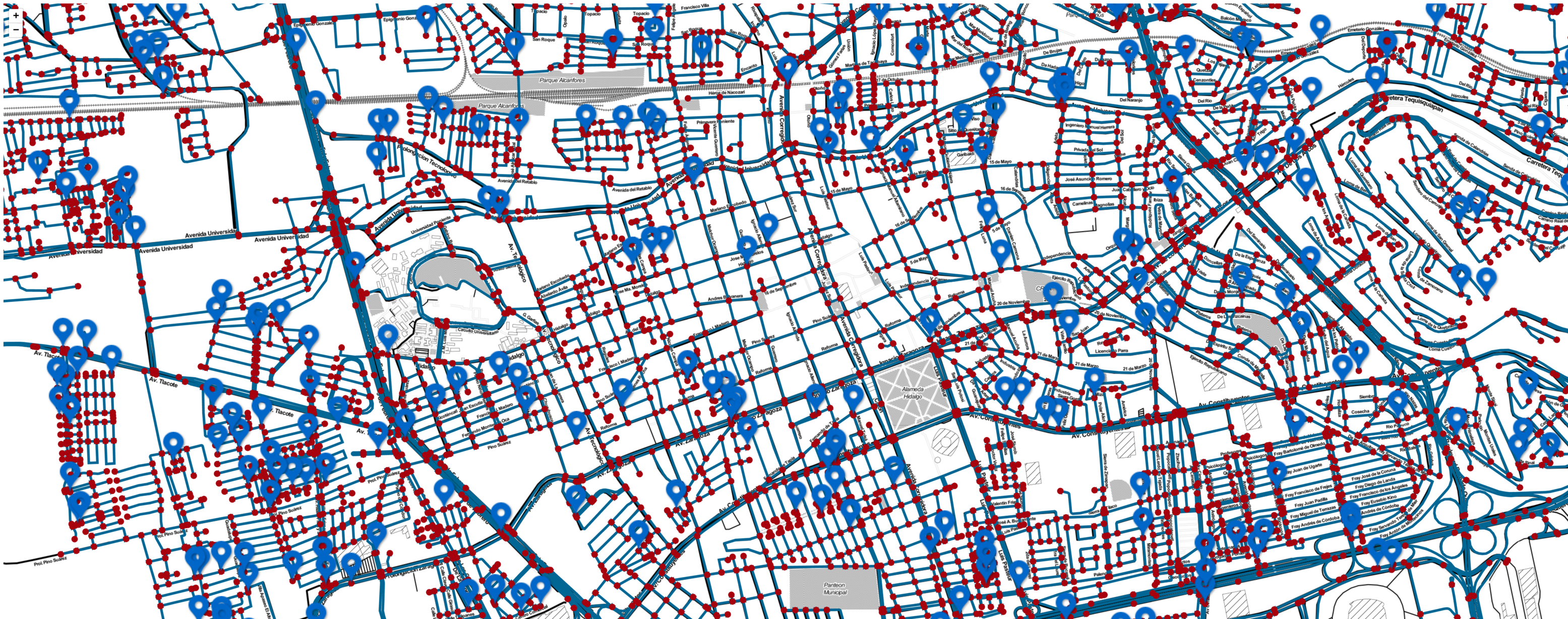}} 

\caption{Screenshots of a Dash Sylvereye visualization displaying the street network of Queretaro City, Mexico, on top of an OSM tilemap layer, at different zoom levels and by showing different visualization layers. The visualized street network has 20,385 nodes and 49,137 edges.}
\label{fig:screenshots} 
\end{figure*}
\fi

Requirements R3, R4, and R5 have been previously identified by the authors in \cite{Han2021} as relevant for high-performance complex graph visualization after interviewing experts in the field and reviewing a series of state-of-the-art tools. Authors in \cite{Huang2015} also acknowledge that, when it comes to studying urban networks with trajectory data, ``the approach needs to handle a large number of city streets and massive trajectory data.'' Regarding requirement R6, authors in \cite{Hadlak2015} note that ``node properties and edge weights play a fundamental role in the field of multivariate network visualization,'' in the context of multifaceted graph visualization.

To address the aforementioned requirements, we developed Dash Sylvereye, a visual analytics library for generating graph-based and interactive visualizations of large street networks and their associated multivariate data. It offers the following solutions to the identified requirements:

\begin{itemize}
\item \textit{R1.} GPU-accelerated rendering of nodes to represent junctions and street road polylines on top of Leaflet.js interactive web maps.
\item \textit{R2.} GPU-accelerated rendering of fully customizable markers. Popups with custom text are supported.
\item \textit{R3.} Nodes, edges, and markers are rendered with WebGL for responsive and ``smooth'' navigation on networks with thousands of elements on any graphics adapter supported by modern web browsers.
\item \textit{R4.} Nodes, edges, and markers styles are customizable: color, size/width, transparency, and visibility. In the case of markers, the user can also customize the marker icon by providing a custom image. Color scales are supported and computed by the library.
\item \textit{R5.} Dash callback triggering when clicking individual nodes, edges, and markers. The user can also listen for changes in the map zoom level and any other property of the visualization component.
\item \textit{R6.} Individual nodes, edges, and markers can be associated with any arbitrary data. Functions are provided to load not only street network topologies but the data associated with them from OSM. The library uses simple list-of-dictionaries data structures for easy loading of networks from any other source.
\item \textit{R7.} The library is implemented as a component of the widely-used Dash framework. This enables Dash Sylvereye visualizations to be natively embedded into custom dashboards. In this way, Dash Sylvereye allows for the display of multivariate data with the help of the plotting components available in Dash, such as bar plots, line plots, and scatter plots. Dash integration also allows the user to coordinate Dash Sylvereye visualizations with a variety of Dash UI elements such as buttons, sliders, dropdown lists, etc.
\end{itemize}

A fully working and complete version of Dash Sylvereye for the Python programming language has been implemented. The following sections describe its design and implementation.

\subsection{Design}

\subsubsection{Layers}

Dash Sylvereye is implemented as a Dash framework component. A Dash Sylvereye visualization is made of four layers: 

\begin{enumerate}
    \item \textit{Tile layer.} Displays a zoomable and pannable web map generated by joining dozens of individually requested images in real time. Dash Sylvereye is built on top of Leaflet.js, enabling the user to select the tilemap provider of his/her preference (e.g. OSM and Mapbox).
    \item \textit{Edge layer.} Displays a clickable polygon for each edge in the street network. It also displays a direction arrow sprite for each edge. It can display edges with different widths, transparency, and color.
    \item \textit{Node layer.} Displays a clickable sprite for each node in the street network. It can display nodes with different sizes, transparency, and color.
    \item \textit{Marker layer.} Displays a clickable sprite for each marker. It can display markers with different sizes, transparency, color, and icon.
\end{enumerate}

\ifheavyimages
\begin{figure*}
\centering
\includegraphics[scale=1.2]{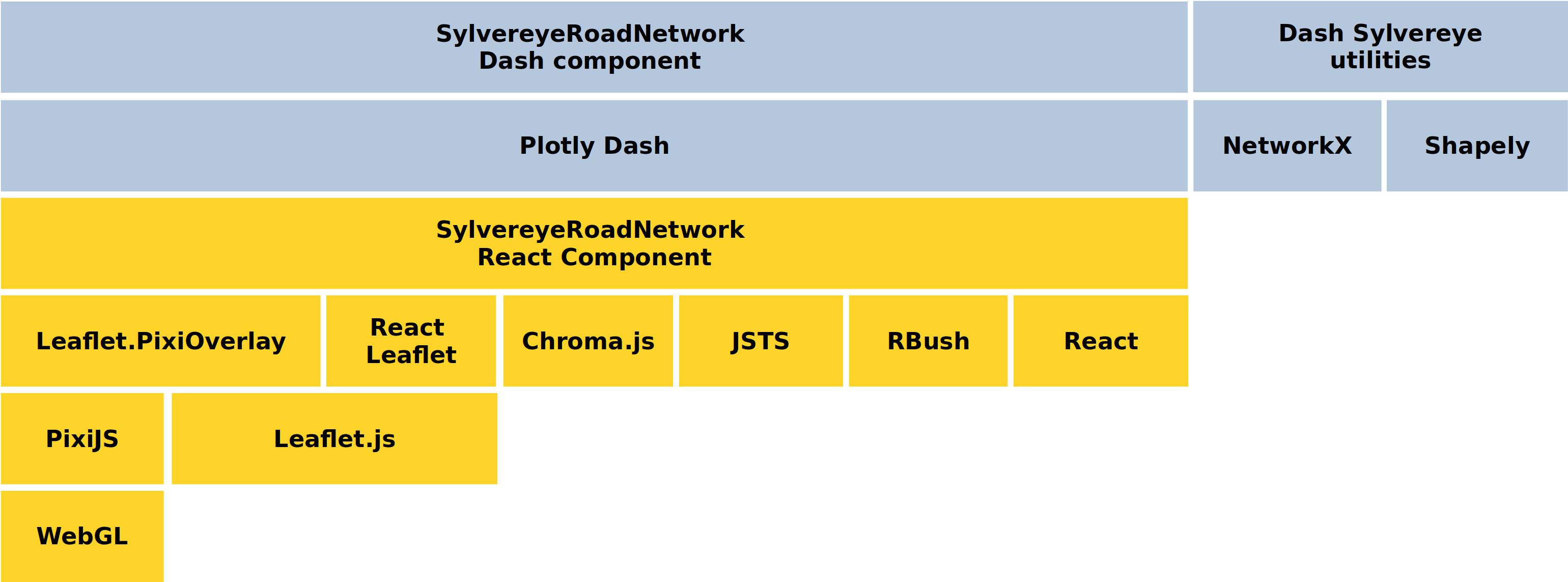} 

\caption{Stack of the main libraries used to build the Dash Sylvereye library. The diagram can be read in the top-down direction as follows: the library in a given layer uses the libraries in the layer located immediately below. The boxes in blue are Python libraries. Boxes in yellow are JavaScript libraries.}
\label{fig:architecture}
\end{figure*}
\fi

Fig. \ref{fig:screenshots} shows screenshots of a Dash Sylvereye visualization displaying the street network of Queretaro City, Mexico, on top of an OSM tilemap, at different zoom levels and with different layers activated. The user can navigate through the visualization by panning and zooming it.

\subsubsection{Data loading}

Dash Sylvereye provides various convenient routines for loading street networks out of NetworkX graphs and GraphML files generated by the OSMnx library. In this way, the user can retrieve the street network of any city from OSM for visualization with a simple query in a single line of code.

\subsubsection{Styling}

The style of individual nodes, edges, and markers is customizable, allowing for the programmatic manipulation of colors, sizes, transparency, and visibility of individual graph elements. The user can also instruct Dash Sylvereye to automatically scale the size, color, and transparency based on values found in the street network's data. When using this coloring option, the user can decide whether to use a predefined or a custom color scale. Markers can show custom popup messages and the default marker's icon can be replaced by a custom SVG image. 

\subsubsection{Interactivity and coordination}

As previously mentioned, nodes, edges, and markers are clickable, allowing for the definition of custom behavior at the user interaction. In addition, the callback architecture of the Dash framework enables the interaction between a Dash Sylvereye visualization and other Dash components. More specifically, any of the visual styles, the street network data, and the street network itself can be updated at runtime as a reaction to events emitted by other Dash components, such as time sliders and buttons. In this way, for example, the transparency and color of edges can be scaled to their vehicle count at different points in time selected via a Dash slider. This specific example would add support for the time dimension to Dash Sylvereye, allowing for the visualization of dynamic events in the street network.
 
\subsubsection{Software and GPU acceleration}

Dash Sylvereye exploits synchronous software (CPU) and GPU acceleration for displaying a large graph on a tiled web map as follows.

Dash Sylvereye uses PixiJS to draw the network. In a regular multimedia application (e.g. a videogame) written with PixiJS the main job of the GPU is to draw each frame efficiently to give the feeling of a smooth animation. In Dash Sylvereye the drawing of a road network and the markers represents the redrawing of a single frame of such an animation. Thus, the GPU function is to render that animation frame as fast as possible. To that end, Dash Sylvereye uses Leaflet.PixiOverlay which allows to draw over a Leaflet.js overlay with PixiJS, which in turn uses WebGL for GPU-accelerated drawing of thousands of objects.

Due to the use of Leaflet.PixiOverlay, the GPU acceleration is involved when: (1) drawing the road network and markers for the first time and (2) redrawing the road network and markers after the user has interacted with it. More specifically, drawing/redrawing (and thus GPU acceleration) is triggered at three specific times:

\begin{enumerate}
    \item At startup (first draw).
    \item After panning the map, i.e. after releasing the left mouse button after panning.
    \item After panning the map, i.e. after zooming in or zooming out.
\end{enumerate}

The rest of the time (i.e. during the panning and zooming animations) the CPU handles the drawing work with software acceleration since the road network was already rendered.

\subsection{Software stack}
 
The Dash Sylvereye library is built on top of the following open-source JavaScript and Python libraries:

JavaScript libraries:
\begin{itemize}
\item \textit{PixiJS}\footnote{https://www.pixijs.com/}: Cross-device 2D rendering library accelerated by WebGL for creating interactive graphics on web browsers. It acts as an abstraction layer for the WebGL API.
\item \textit{Leaflet.js}\footnote{https://leafletjs.com/}: Mapping library for rendering interactive tiled web maps hosted on public servers with (optional) tiled overlays. Supports HTML5 and CSS3. It can create interactive layers.
\item \textit{Leaflet.PixiOverlay}\footnote{https://github.com/manubb/Leaflet.PixiOverlay}: Overlay class for Leaflet.js for WebGL-accelerated drawing on top of tiled web maps using PixiJS.
\item \textit{Chroma.js}\footnote{https://gka.github.io/chroma.js/}: Library for computing color conversions and color scales in the web browser. 
\item \textit{JSTS}\footnote{https://bjornharrtell.github.io/jsts/}: Library of spatial predicates and functions for processing geometries in web browsers. It is a JavaScript port of the JTS Java library.
\item \textit{RBush}\footnote{https://github.com/mourner/rbush}: Library for 2D spatial indexing of points and rectangles in web browsers. It is built around a custom R-tree data structure with bulk insertion support.
\item \textit{React.js}: Component-driven front-end library for building UI components maintained by Facebook.
\item \textit{React Leaflet}\footnote{https://react-leaflet.js.org/}: Bindings between React.js and Leaflet.js. Exposes Leaflet.js layers as React components. 
\end{itemize}

Python libraries:
\begin{itemize}
\item \textit{Plotly Dash}: User interface library for creating data-driven web applications around dashboard visualizations entirely in Python.
\item \textit{NetworkX}\footnote{https://networkx.org/}: Social Network Analysis library for network reading, creation, generation, manipulation, measuring, and visualization.
\item \textit{Shapely}\footnote{https://github.com/Toblerity/Shapely}: Library for manipulating geometric objects in the Cartesian plane.
\end{itemize}

Fig. \ref{fig:architecture} shows the library stack used to develop Dash Sylvereye. Leaflet.js provides a layer of tiled web maps as well as zooming and panning capabilities, whereas the PixiJS library provides WebGL-powered street network drawing primitives (polygons and sprites). This is done by using Leaflet.PixiOverlay which provides a Leaflet.js overlay where PixiJS can draw. 

Dash Sylvereye also makes use of other third-party JavaScript libraries, such as JSTS for defining edge-hit polygons, RBush for efficiently finding edge-hit polygons that have been clicked by the user, and Chroma.js for computing color scales for edges, nodes, and markers. 

React Leaflet is used to bring everything together: the Leaflet.js map, the tilemap layer, and the road network visualization layer that exploits Leaflet.PixiOverlay. All these elements are encapsulated into the \textit{SylvereyeRoadNetwork} React Component. The React component is then wrapped to produce the \textit{SylvereyeRoadNetwork} Dash component by using the toolchain provided by Dash.
 
On the Python side, Dash Sylvereye network loading routines make use of NetworkX and Shapely, enabling Dash Sylvereye to import street networks from the OSM project via OSMnx or from OSMnx-generated GraphML files.


\section{Usage example}\label{section:usage-example}

\ifheavyimages
\begin{figure*}
\includegraphics[scale=1.0]{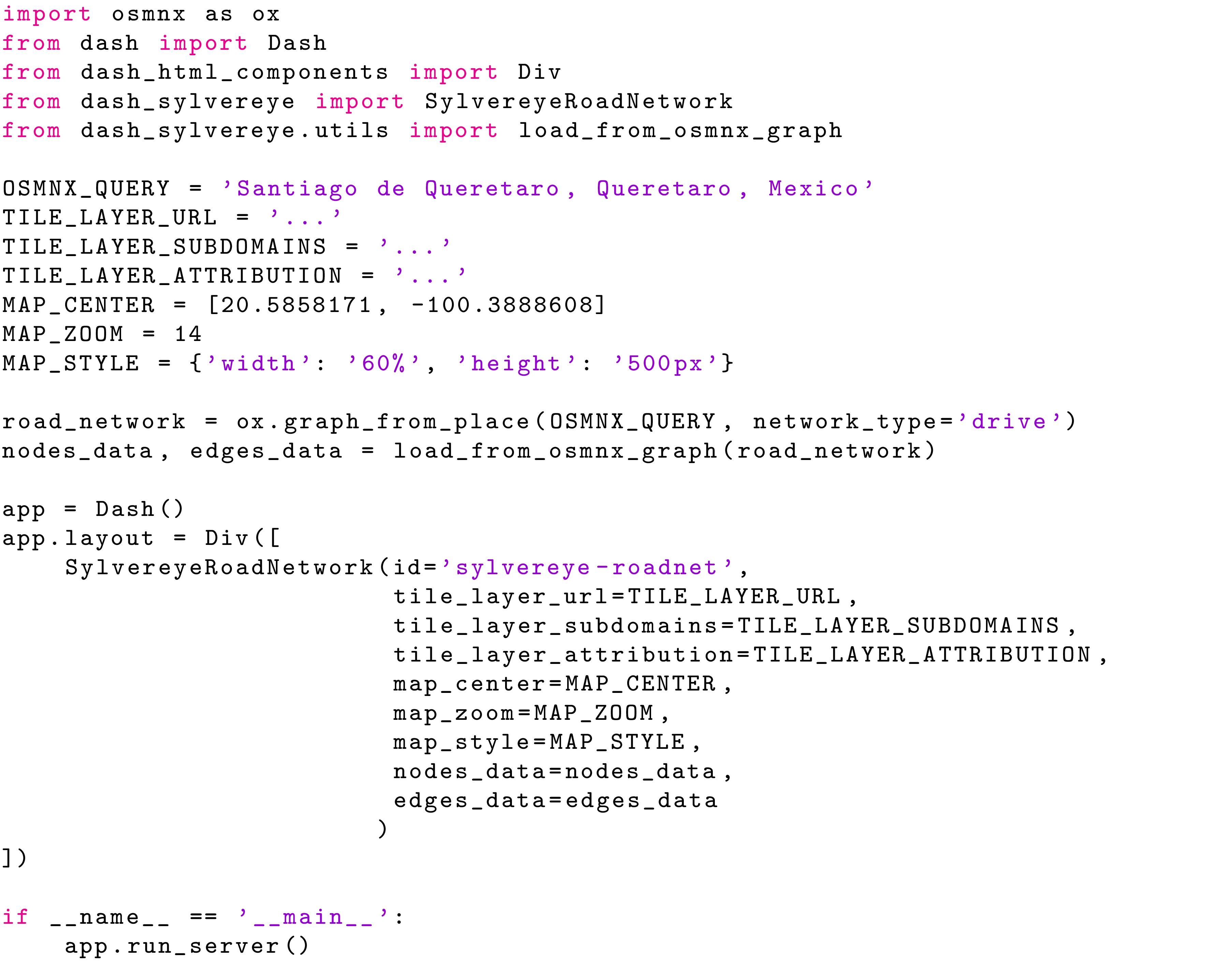} 

\caption{Example showing how to embed a SylvereyeRoadNetwork component in a minimal Dash dashboard to display a street network obtained with OSMnx.}
\label{fig:code-example-1}
\end{figure*}

\begin{figure*}
\includegraphics[scale=1.0]{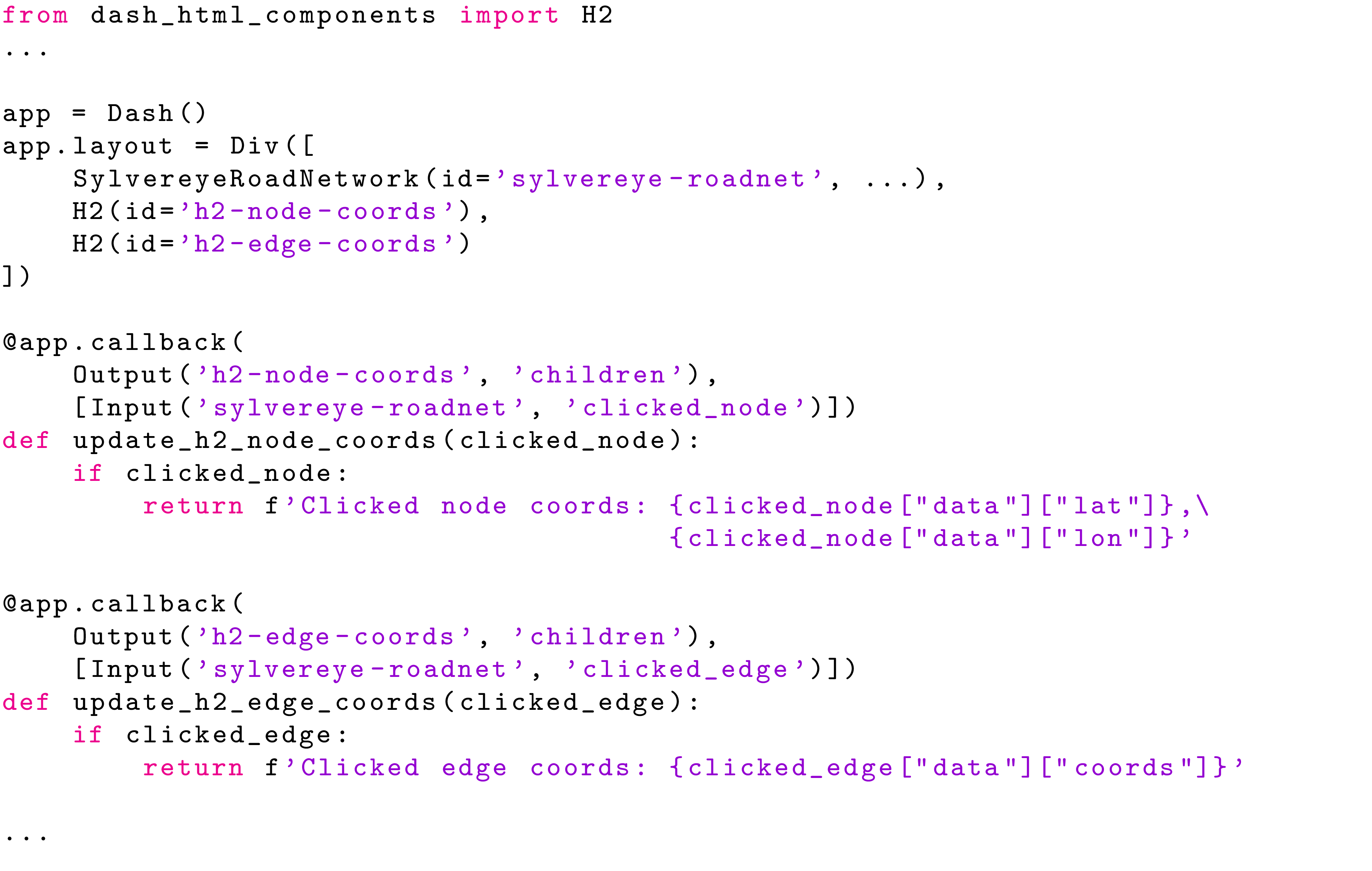} 

\caption{Example showing how to use Dash callbacks to react to mouse clicks on the street network’s nodes and edges.}
\label{fig:code-example-2}
\end{figure*}

\begin{figure*}
\includegraphics[scale=1.0]{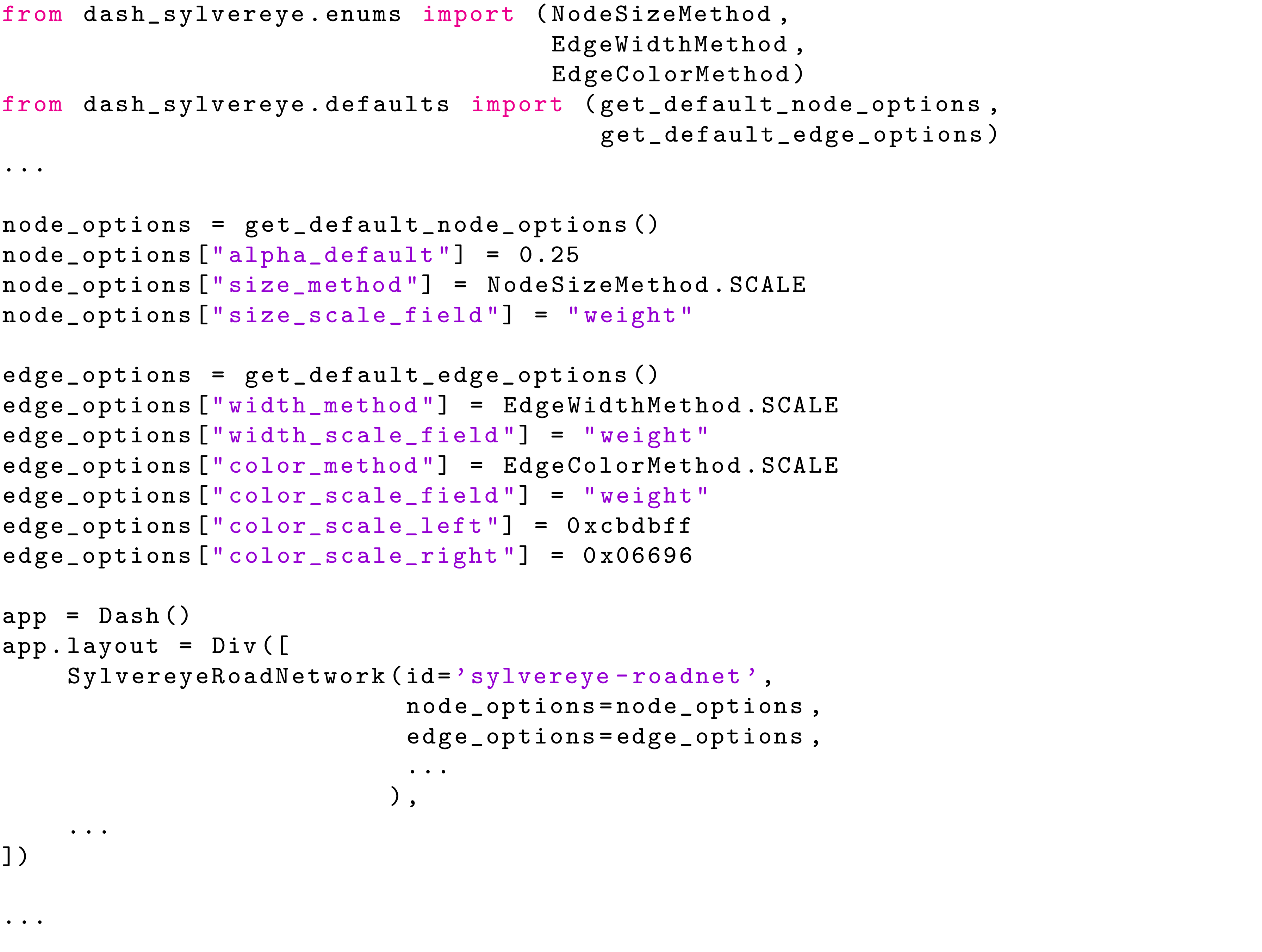} 

\caption{Example showing how to configure the visual styles of nodes (transparency and size) and edges (width and color scale) in a Dash Sylvereye visualization.}
\label{fig:code-example-3}
\end{figure*}

\begin{figure*}
\centering
\includegraphics[scale=1.35]{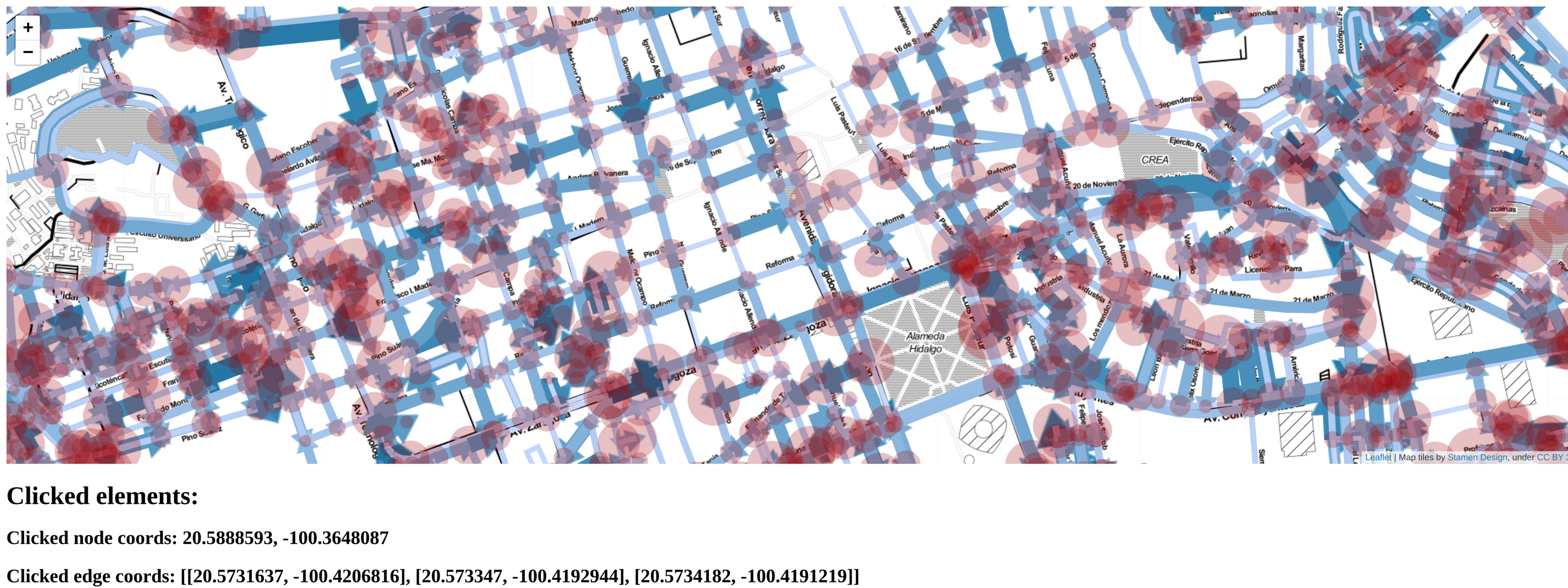} 

\caption{Screenshot of the resulting dashboard example after putting together the code snippets shown in Figs. \ref{fig:code-example-1}-\ref{fig:code-example-3}.}
\label{fig:example-screenshot}
\end{figure*}
\fi

\begin{table*}
\caption{The SylvereyeRoadNetwork Dash component supports an array of properties classified as follows: 1) data properties, 2) (style) option properties, 3) show/hide properties, map properties, 4) tile layer properties, and 5) callback properties. All but the callback properties are provided and updated by the user to set up and tune the street network visualization. Callback properties, on the other hand, are updated by Dash as a reaction to user-click interaction.}
\label{table:sylvereye-parameters}
\begin{centering}
\begin{tabular}{|p{3.7cm}|p{1.3cm}|p{5.5cm}|p{5.5cm}|}
\hline 
Property & Category & Brief description & Observations\tabularnewline
\hline 
\verb|nodes_data| & Data & List of the road network's nodes & Nodes can hold arbitrary data in the 'data' field\tabularnewline
\verb|edges_data| & Data & List of the road network's edges & Edges can hold arbitrary data in the 'data' field\tabularnewline
\verb|markers_data| & Data & List of map markers & Markers can hold arbitrary data in the 'data' field\tabularnewline
\verb|node_options| & Options & Visual options dictionary for nodes & -\tabularnewline
\verb|edge_options| & Options & Visual options dictionary for edges & -\tabularnewline
\verb|marker_options| & Options & Visual options dictionary for markers & -\tabularnewline
\verb|show_nodes| & Show/hide & If false, all nodes will be hidden & Hidden nodes will cease to be interactive\tabularnewline
\verb|show_edges| & Show/hide & If false, all edges will be hidden & Hidden edges will cease to be interactive\tabularnewline
\verb|show_arrows| & Show/hide & If false, all direction arrows will be hidden & -\tabularnewline
\verb|show_markers| & Show/hide & If false, all markers will be hidden & Hidden markers will cease to be interactive\tabularnewline
\verb|map_center| & Map & Map center coordinates & In (latitude, longitude) format\tabularnewline
\verb|map_zoom| & Map & Map zoom level & As specified by Leaflet.js\tabularnewline
\verb|map_min_zoom| & Map & Minimum allowed map level & As specified by Leaflet.js\tabularnewline
\verb|map_max_zoom| & Map & Maximum allowed map level & As specified by Leaflet.js\tabularnewline
\verb|map_style| & Map & Map CSS styles & Provided in dictionary form: \{'style': 'value'\}\tabularnewline
\verb|tile_layer_url| & Tile layer & Tile layer URL template & As specified by Leaflet.js\tabularnewline
\verb|tile_layer_subdomains| & Tile layer & Tile layer attribution HTML text & As specified by Leaflet.js\tabularnewline
\verb|tile_layer_attribution| & Tile layer & Tile layer subdomains & As specified by Leaflet.js\tabularnewline
\verb|tile_layer_opacity| & Tile layer & Tile layer opacity & A value between 0 and 1\tabularnewline
\verb|clicked_node| & Callback & Used to invoke a callback when a node is clicked & Data will be available as the property's value\tabularnewline
\verb|clicked_edge| & Callback & Used to invoke a callback when an edge is clicked & Data will be available as the property's value\tabularnewline
\verb|clicked_marker| & Callback & Used to invoke a callback when a marker is clicked & Data will be available as the property's value\tabularnewline
\hline 
\end{tabular}
\par\end{centering}
\end{table*}

\begin{table*}
\caption{Style methods available for nodes, edges, and markers. For example, there are three color methods for nodes: NodeColorMethod.DEFAULT, NodeColorMethod.SCALE, and NodeColorMethod.CUSTOM. DEAFAULT methods use the predefined settings provided by Dash Sylvereye, which can be customized. SCALE methods, on the other hand, scale visual style values in proportion to a weight field. CUSTOM methods allow styling based on the data associated with individual nodes, edges, and markers. For the case of visibility methods, ALWAYS instructs Dash Sylvereye to turn the visibility of all elements on. Some other style methods are more specific to a given kind of element, such as the ORIGINAL method for marker icons, which makes Dash Sylvereye use the original color of the SVG image specified as an icon.} 
\label{table:visual-options}
\begin{centering}
\begin{centering}
\begin{tabular}{|p{1.5cm}|p{4.5cm}|p{4.5cm}|p{5cm}|}
\hline 
Style & Node methods & Edges methods & Marker methods\tabularnewline
\hline 
Color & \verb|DEAFAULT|, \verb|SCALE|, \verb|CUSTOM| & \verb|DEAFAULT|, \verb|SCALE|, \verb|CUSTOM| & \verb|DEAFAULT|, \verb|SCALE|, \verb|CUSTOM|, \verb|ORIGINAL|\tabularnewline
Size & \verb|DEAFAULT|, \verb|SCALE|, \verb|CUSTOM| & N/A & \verb|DEAFAULT|, \verb|SCALE|, \verb|CUSTOM|\tabularnewline
Alpha & \verb|DEAFAULT|, \verb|SCALE|, \verb|CUSTOM| & \verb|DEAFAULT|, \verb|SCALE|, \verb|CUSTOM| & \verb|DEAFAULT|, \verb|SCALE|, \verb|CUSTOM|\tabularnewline
Visibility & \verb|ALWAYS|, \verb|CUSTOM| & \verb|ALWAYS|, \verb|CUSTOM| & \verb|ALWAYS|, \verb|CUSTOM|\tabularnewline
Width & N/A & \verb|DEAFAULT|, \verb|SCALE|, \verb|CUSTOM| & N/A\tabularnewline
Icon & N/A & N/A & \verb|DEFAULT|, \verb|CUSTOM|\tabularnewline
\hline 
\end{tabular}
\par\end{centering}
\par\end{centering}
\vspace{0.2cm}
\end{table*}

This section presents a simple usage example to illustrate what programming with Dash Sylvereye looks like. The example is separated into three parts, namely initialization, interactivity, and styling.

In Fig. \ref{fig:code-example-1}, the street network of Queretaro City is retrieved from OSM with the OSMnx library and then transformed to Dash Sylvereye's list-of-dictionaries data structure by using the utility function \verb|load_from_osmnx_graph()|. Dash Sylvereye also provides the function \verb|load_from_osmnx_graphml()| to load a street network from a graph file in GraphML format generated by OSMnx.

To insert a street network visualization in a Dash dashboard, the programmer only has to insert an instance of the class \verb|SylvereyeRoadNetwork| in the dashboard application layout. The street network topology (and data) is provided via the \verb|nodes_data| and \verb|edges_data| parameters. Apart from the road network, the user can provide information about the map and the web tile layer by using an interface similar to that of Leaflet.js. 

Table \ref{table:sylvereye-parameters} shows the list of currently supported properties in the \verb|SylvereyeRoadNetwork| class. These properties allow the user to set up the tilemap (e.g. tilemap provider and attribution), the road network data (e.g. nodes and edges), node/edge/marker style options, the layer visibility, and the map itself (e.g. zoom level and center). Recall that any of these parameters can be updated at runtime, triggering the automatic update of the visualization when changed. For example, if the user wants to update the street network topology, it is enough to update the \verb|nodes_data| and the \verb|edges_data| parameters in a callback.

Fig. \ref{fig:code-example-2} shows an example on the use of callbacks for reacting to user interaction by using the \verb|clicked_node| and \verb|clicked_edge| callback parameters listed in Table \ref{table:sylvereye-parameters}. Every time the user clicks a node, a callback provided by the programmer is triggered to update an \verb|H2| Dash label component with the node's coordinates. Likewise, every time the user clicks an edge the provided callback is triggered to update an \verb|H2| label component label with the edge's polyline coordinates.

The programmer can fine-tune the visuals of the street network visualization on an element-by-element basis by filling option dictionaries available for nodes, edges, and markers. Table \ref{table:visual-options} lists the currently supported style option methods. The user only needs to: 1) get an options dictionary pre-filled with default settings, 2) customize the options dictionary by selecting and setting up one or more visual option methods listed in Table \ref{table:visual-options}, and 3) pass the dictionary to the Dash Sylvereye component. Again, if the user passes an updated options dictionary to Dash Sylvereye at runtime, the visualization will update accordingly in an automatic fashion. 

In Fig. \ref{fig:code-example-3}, the transparency level (alpha) of all nodes is set to 0.25 to make them translucent. Also, the size method is set to \verb|NodeSizeMethod.SCALE| in order to set the diameter of all nodes in proportion to their weight. As for the visuals of edges, both the edge width and edge color methods are also set to \verb|EdgeWidthMethod.SCALE| and \verb|EdgeColorMethod.SCALE|, respectively, in order to be scaled in proportion to edge weights. Fig. \ref{fig:example-screenshot} shows the resulting web dashboard when putting together the code provided in Figs. \ref{fig:code-example-1}-\ref{fig:code-example-3}.

 
\begin{table*}
\caption{Details of the OSM street networks and the tilemap configuration used for the animation performance assessment.}
\label{table:street-networks}
\begin{centering}
\begin{tabular}{|M{2.0cm}|M{5.0cm}|M{1.5cm}|M{1.5cm}|M{2.0cm}|M{1.5cm}|}
\hline 
City name & OSMnx query string & Number of nodes & Number of edges & Center (lat, lon) & Zoom level\tabularnewline
\hline 
Alameda, US & Alameda, Alameda County, CA, USA & 1,830 & 4,842 & 37.7618235, $-122.2429843$ & 15\tabularnewline
Enschede, NL & Enschede, Overijssel, Netherlands, The Netherlands & 5,337 & 13,587 & 52.2271595, 6.9046205 & 12\tabularnewline
Queretaro, MX & Santiago de Querétaro, Querétaro, México & 20,385 & 49,137 & 20.6025256, $-100.3886302$ & 12\tabularnewline
Beijing, CN & Beijing, China & 63,347 & 153,120 & 39.9116304, 116.4010405 & 9\tabularnewline
\hline  
\end{tabular}
\par\end{centering}
\end{table*}
 
\section{Animation performance assessment}\label{section:animation-performance-assessment}

\begin{figure*}  
\centering
\subfloat[]{\includegraphics[width=0.6\columnwidth]{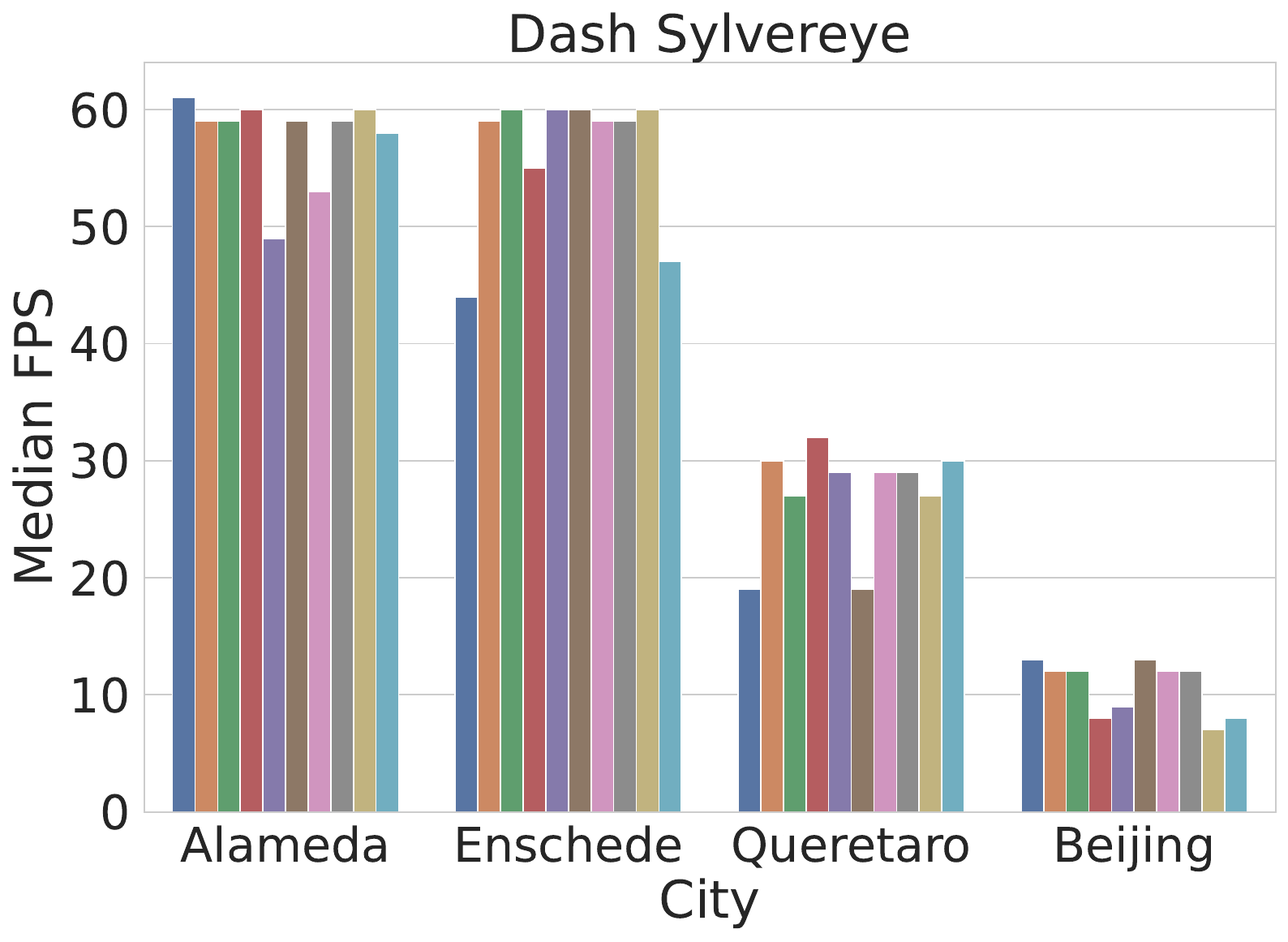}}\hspace{5mm}
\subfloat[]{\includegraphics[width=0.6\columnwidth]{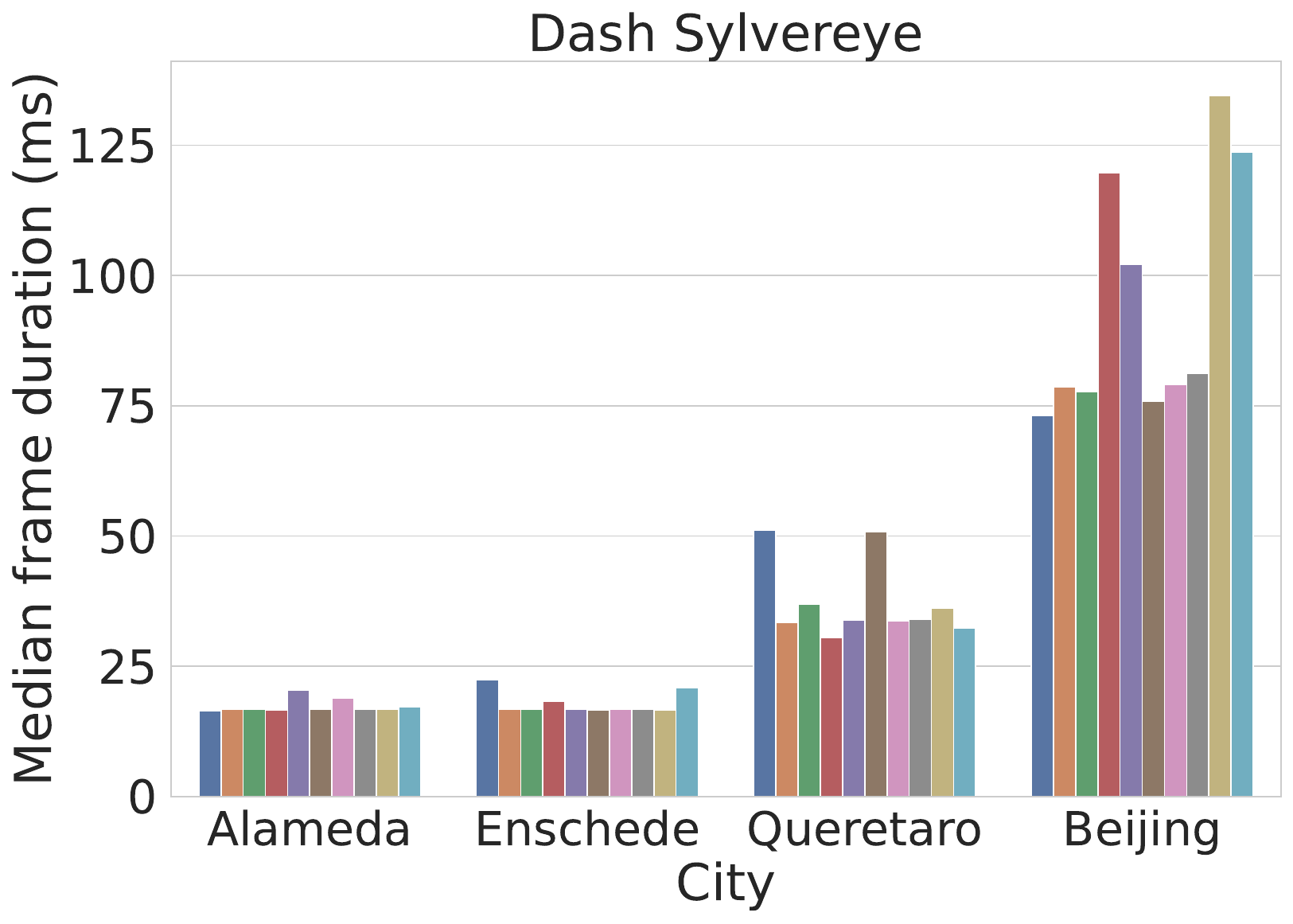}}\hspace{5mm}
\subfloat[]{\includegraphics[width=0.6\columnwidth]{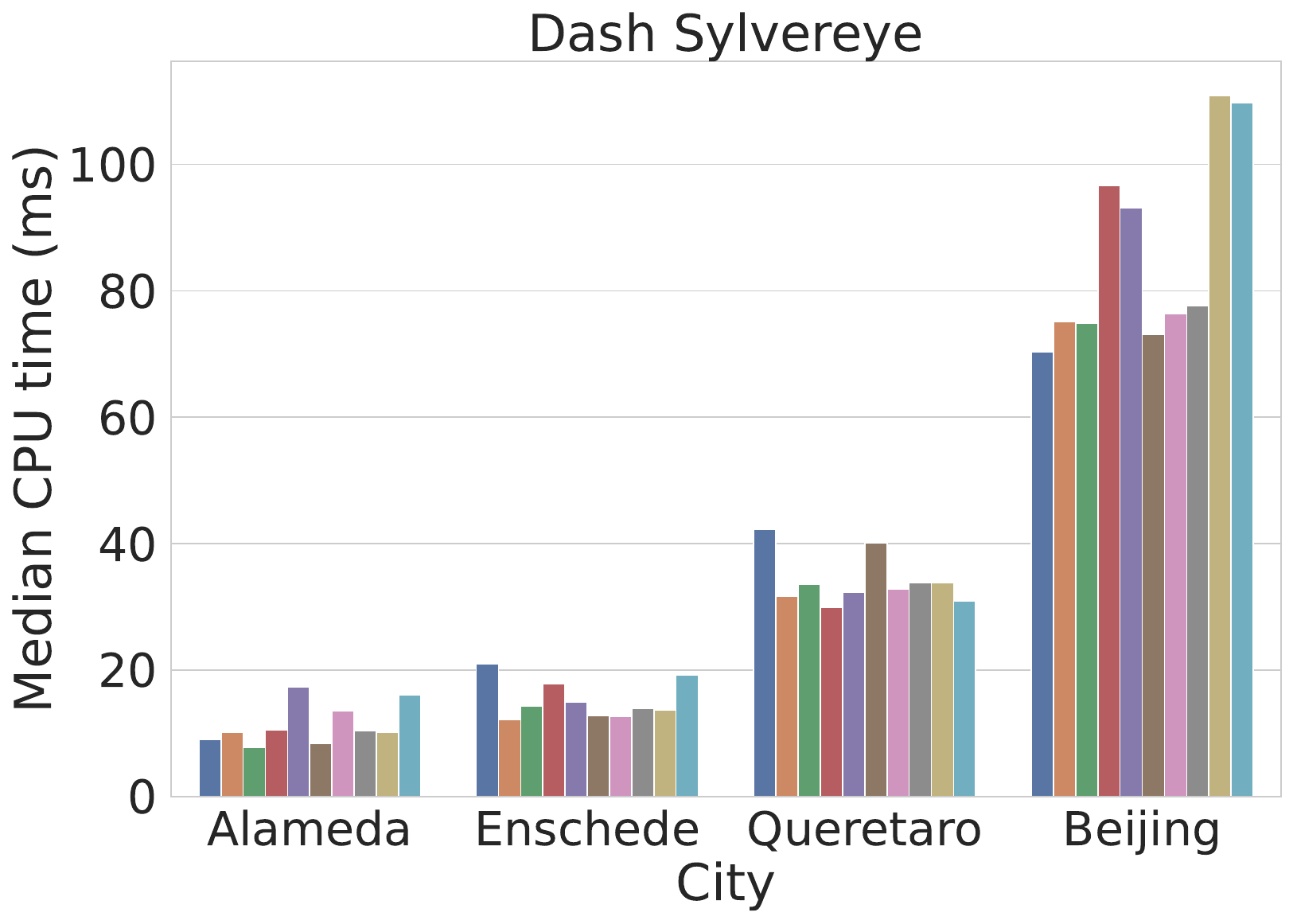}}
\caption{Dash Sylvereye's median frame FPS, duration, and CPU time for each experiment and each city. Cities are sorted from smaller to bigger from left to right.} 
\label{fig:sylvereye-performance-assessment-barplots}
\end{figure*}

\begin{figure*}  
\centering
\subfloat[]{\includegraphics[width=0.6\columnwidth]{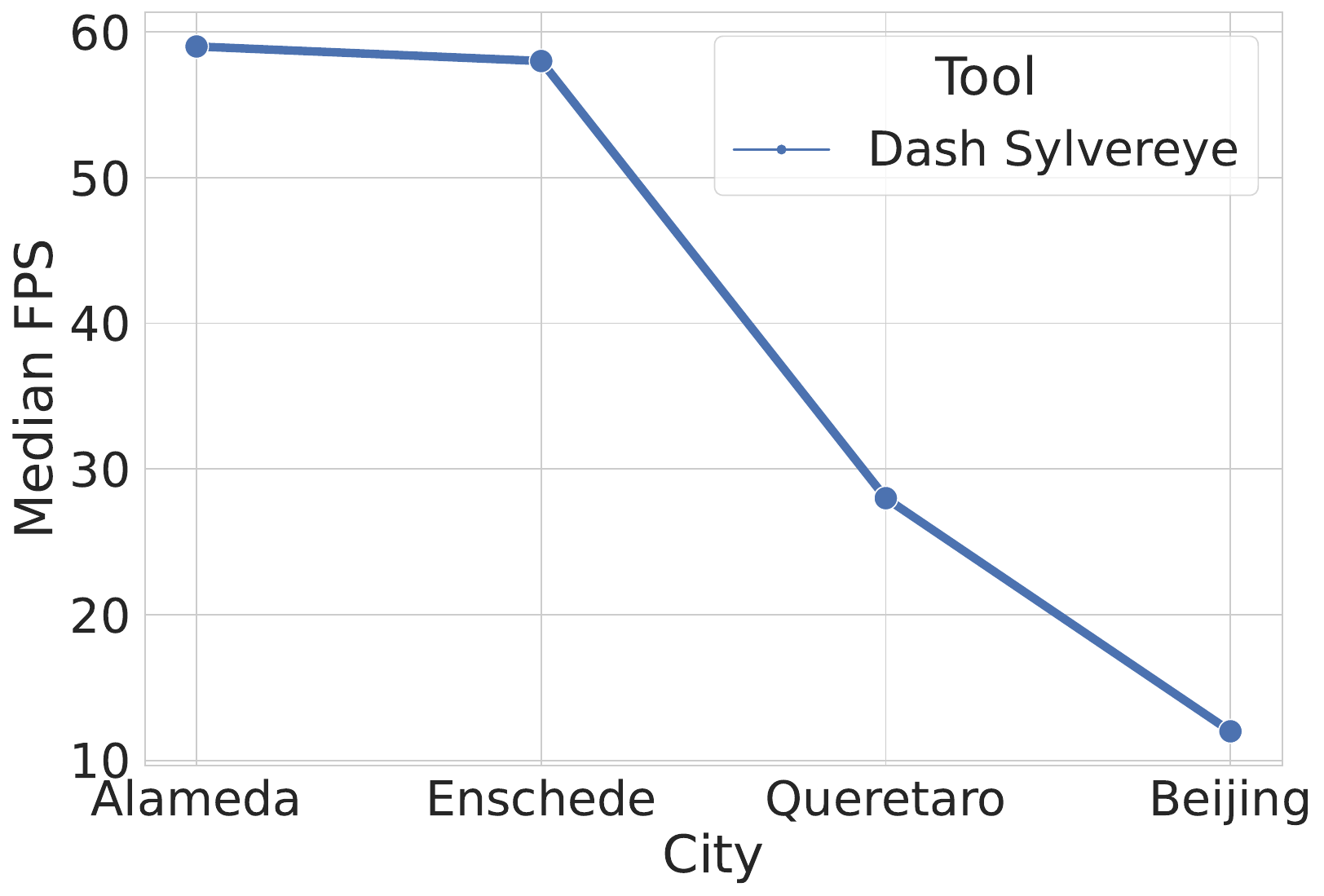}}\hspace{5mm}
\subfloat[]{\includegraphics[width=0.6\columnwidth]{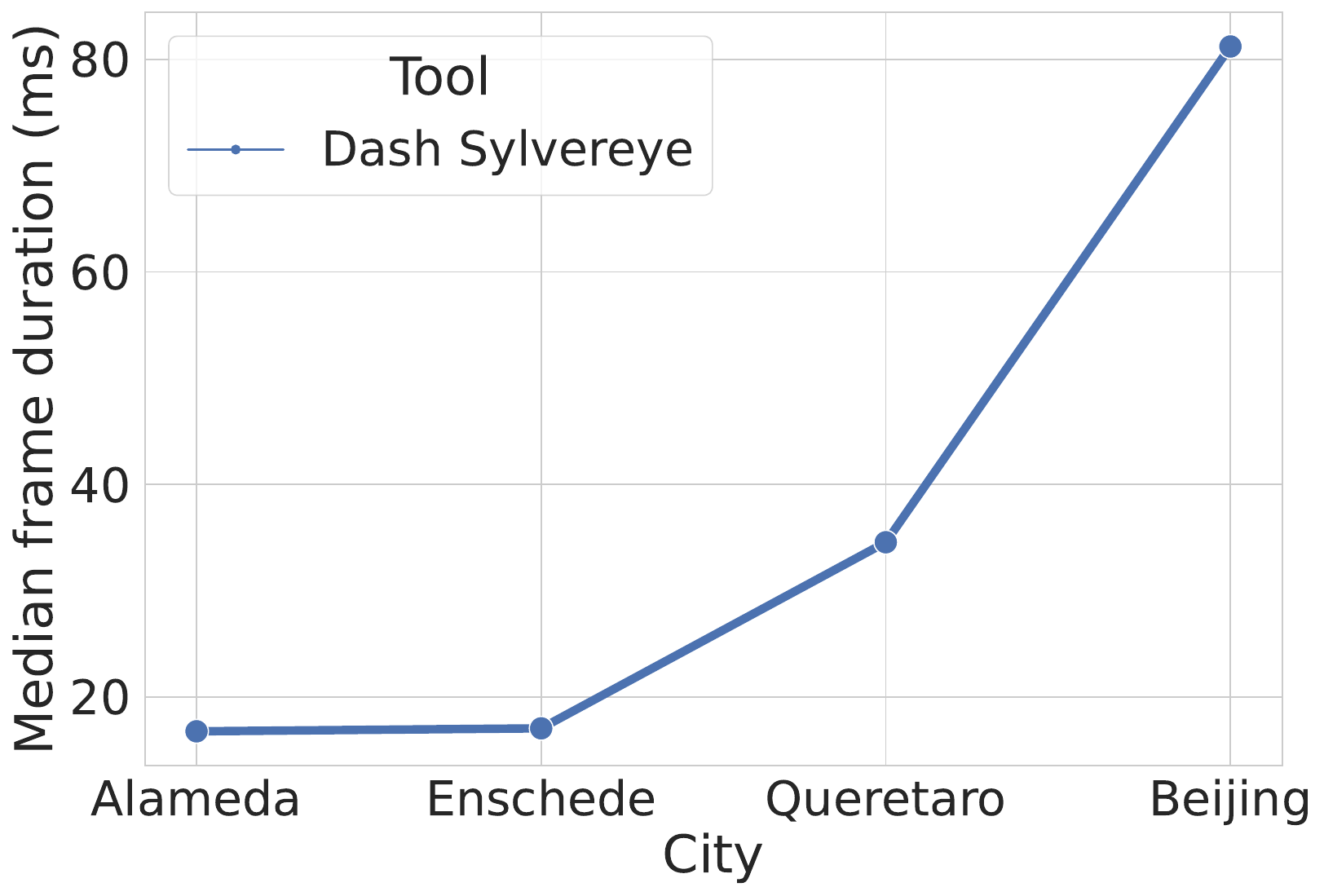}}\hspace{5mm}
\subfloat[]{\includegraphics[width=0.6\columnwidth]{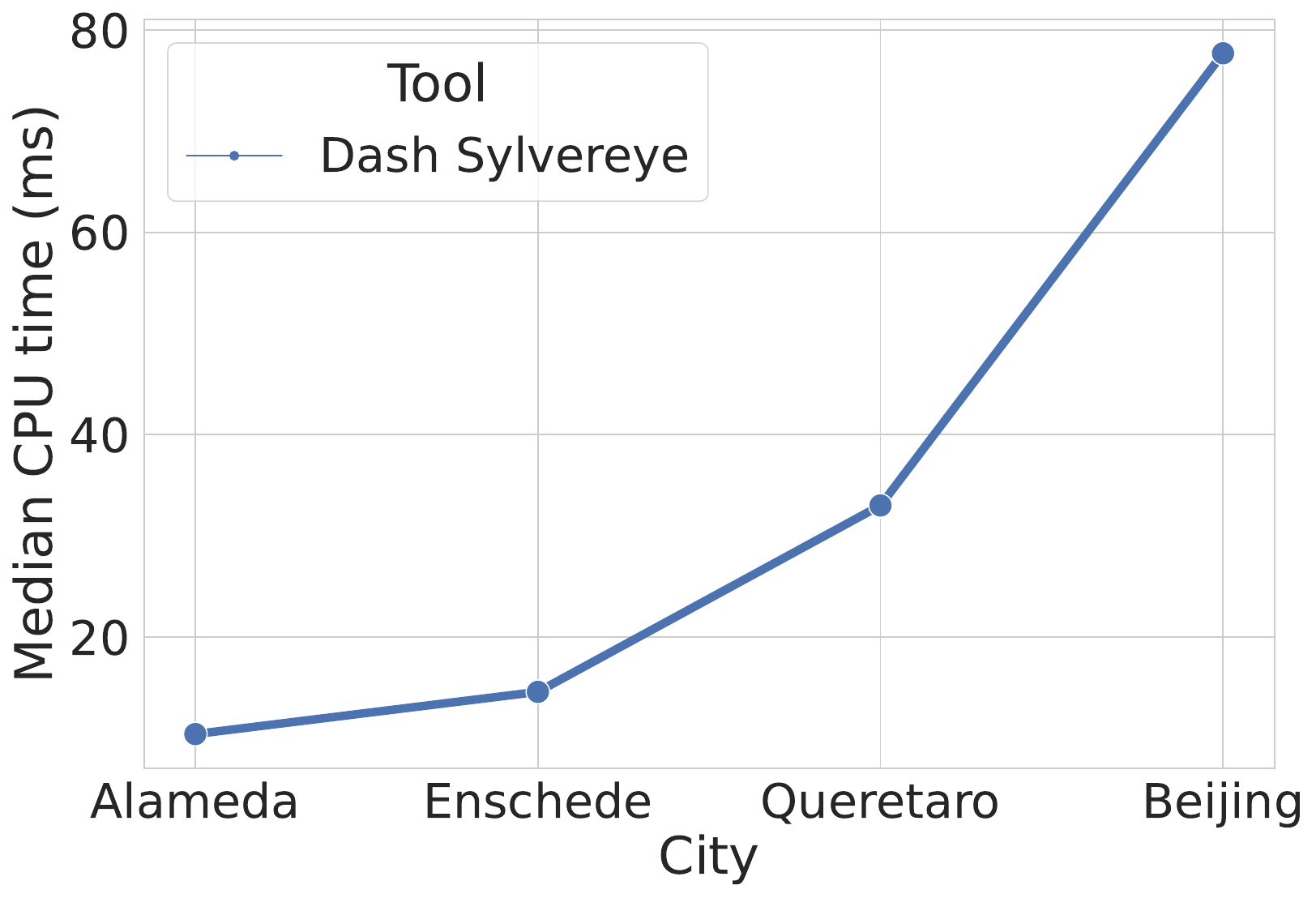}} 
\caption{Dash Sylvereye's median frame FPS, duration, and CPU time when merging all experiments for each city. Cities are sorted from smaller to bigger from left to right.}
\label{fig:sylvereye-performance-assessment-lineplots}
\end{figure*}

We quantitatively assessed how ``responsive'' is to the user interaction with Dash Sylvereye visualizations on a commodity computer for a set of OSM street networks of varying sizes when exploiting software acceleration. 

Panning\footnote{Panning consists in holding the left button of the mouse and moving the mouse to navigate on the map.} of a web map is an important operation since, in our case, it lets the user navigate the road network and explore its elements. We, therefore, assessed how smooth is the panning of a network visualization by measuring the screen refresh rate of a web page in terms of the animation frames per second (FPS). The CPU time and frame duration can offer insights for explaining the observed FPS.
 
We conducted the assessment on a commodity computer with a dual-core AMD A9-9425 processor at 3.1 GHz, with an Integrated AMD Radeon R5 (Stoney Ridge) GPU, and 7.2 GiB in RAM. The computer was running Linux Ubuntu 18.04.4 LTS 64-bits. Note that the processor used in the experiments is a mid-end mobile CPU with an integrated GPU that can be found in budget laptops.

The assessment methodology consisted of two main stages. In the first stage, we retrieved the data of street networks from OSM by running OSMnx with the query strings listed in Table \ref{table:street-networks} for four cities. We used the OSM website to get initial map center coordinates to open the test dashboards and then choose the final map centers and zoom levels. Final map centers and zoom levels were chosen in such a way that the whole street network was visible.
 
In the second stage, we conducted the following experiment for each street network. We used the performance tab of the Chrome DevTools console to record the dashboard while manually panning the whole visualization by performing circular dragging movements. We used Google Chrome v85.0.4183.121.

Next, we manually registered the frame duration, frame FPS, and frame CPU time of 31 recorded animation frames from the Chrome DevTools performance tab to obtain statistically valid results.  We repeated this experiment 10 times for each street network.

Fig. \ref{fig:sylvereye-performance-assessment-barplots} shows the median frame FPS, duration, and CPU time for each experiment and each city. Fig. \ref{fig:sylvereye-performance-assessment-lineplots} shows the median values when merging all experiments for each city. We use the median because it is less sensitive to outliers than the average. Cities are sorted from smaller to larger from left to right.

Figures show that lower FPS values are associated with larger CPU times and frame durations. This might be explained by the fact that the more the CPU has to work the more the duration of an animation frame, negatively impacting the FPS in that animation frame.

From the FPS perspective, figures show that the larger the city the lower the FPS, ranging from around 60 FPS for the Alameda city to around 10 FPS for the Beijing city. Nonetheless, Queretaro city, with 20k nodes and 49k edges, shows an FPS of above 24 FPS, suggesting that Dash Sylvereye can smoothly handle the panning of networks with dozens of thousands of nodes and edges on the experiment machine.

 
\section{Animation performance comparison}\label{section:comparison-to-soa}

We also present a performance comparison between Dash Sylvereye and other state-of-the-art visualization libraries that can render large road networks: Kepler.gl and city-roads. We quantitatively measured and compared the responsiveness to the user interaction of the three tools on a commodity computer for the Alameda, Enschede, Queretaro, and Beijing road networks.

The hardware setup and the two-stage methodology were the same as in Section \ref{section:animation-performance-assessment}. We conducted the 10 experiments for each tool sequentially and continuously in time, without interruptions (no computer reset, no login-logout, etc.) to get numbers as accurate as possible. There were no other apps and tabs other than the web browser was open. The full city map was always visible during the movements in all experiments. All circular movements were performed manually, and clockwise in all experiments. We used the same manual movements with regard to speed and diameter as humanly possible. The center and zoom level in Kepler.gl and Dash Sylvereye were adjusted programmatically whereas for city-roads we had to set the center and zoom level manually. For this experimentation, we used Google Chrome v87.0.4280.88 on Ubuntu 20.04.1 LTS (64-bit).

\begin{figure*}  
\centering
\subfloat[]{\includegraphics[width=0.6\columnwidth]{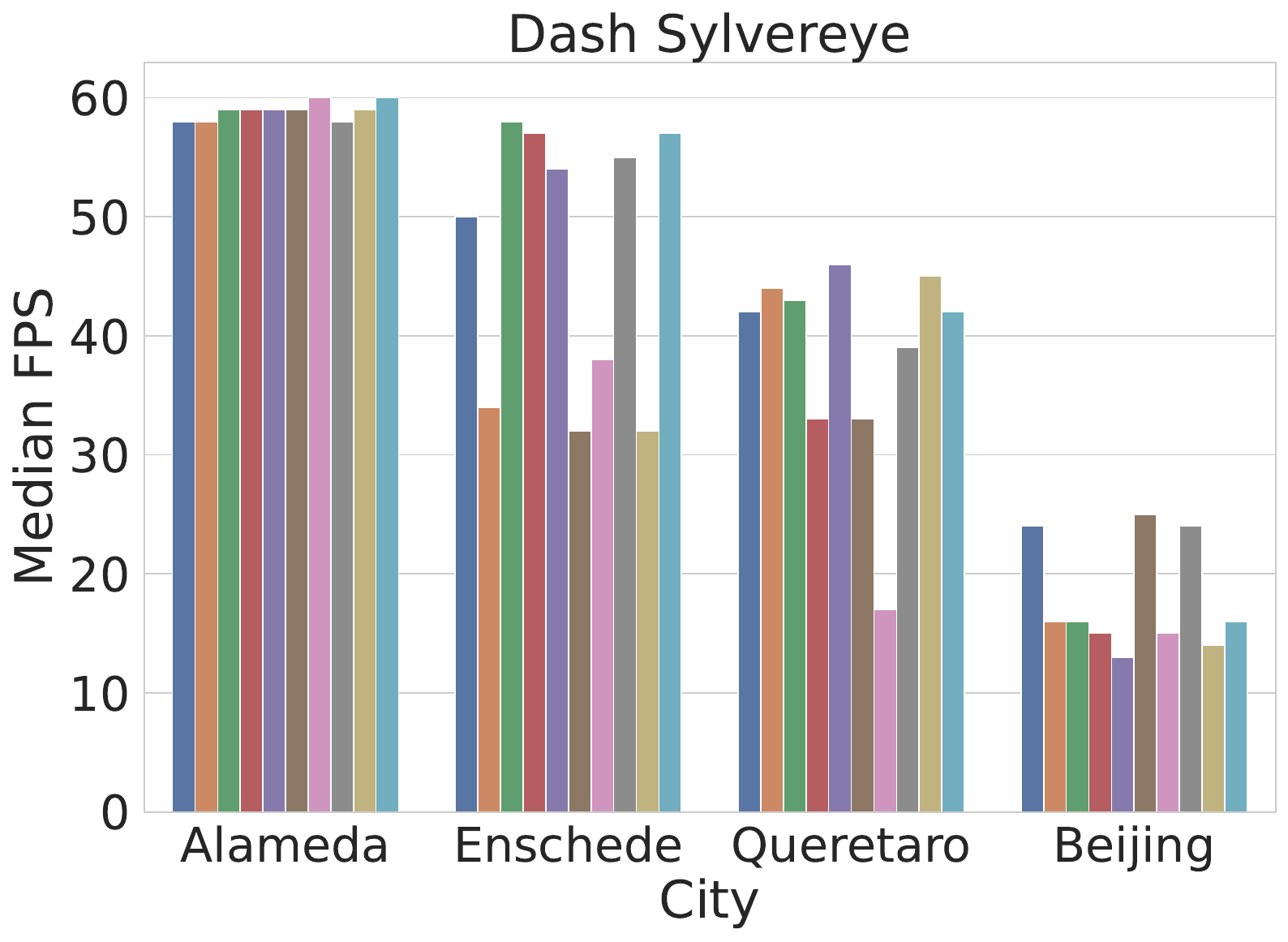}}\hspace{5mm}
\subfloat[]{\includegraphics[width=0.6\columnwidth]{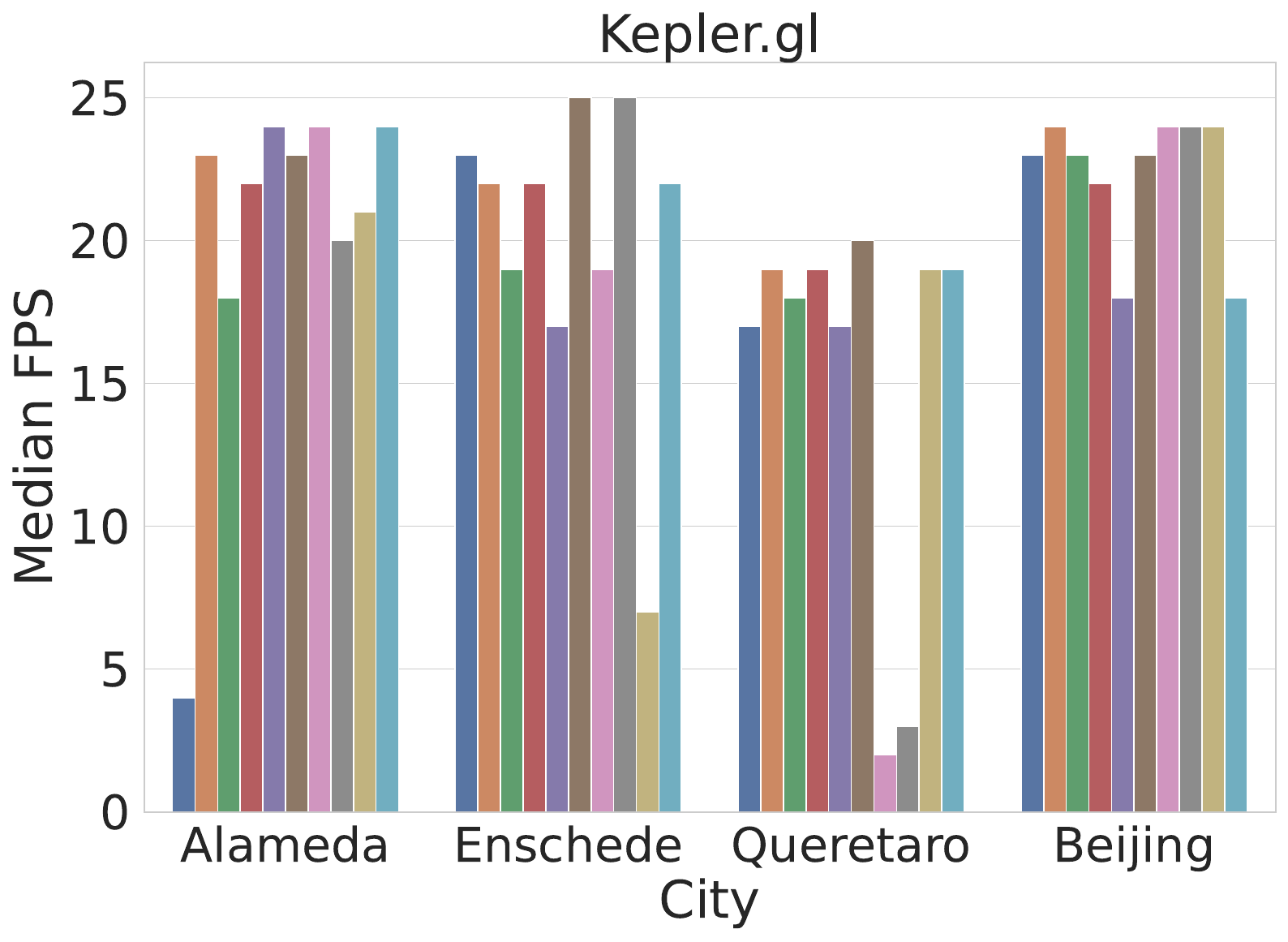}}\hspace{5mm}
\subfloat[]{\includegraphics[width=0.6\columnwidth]{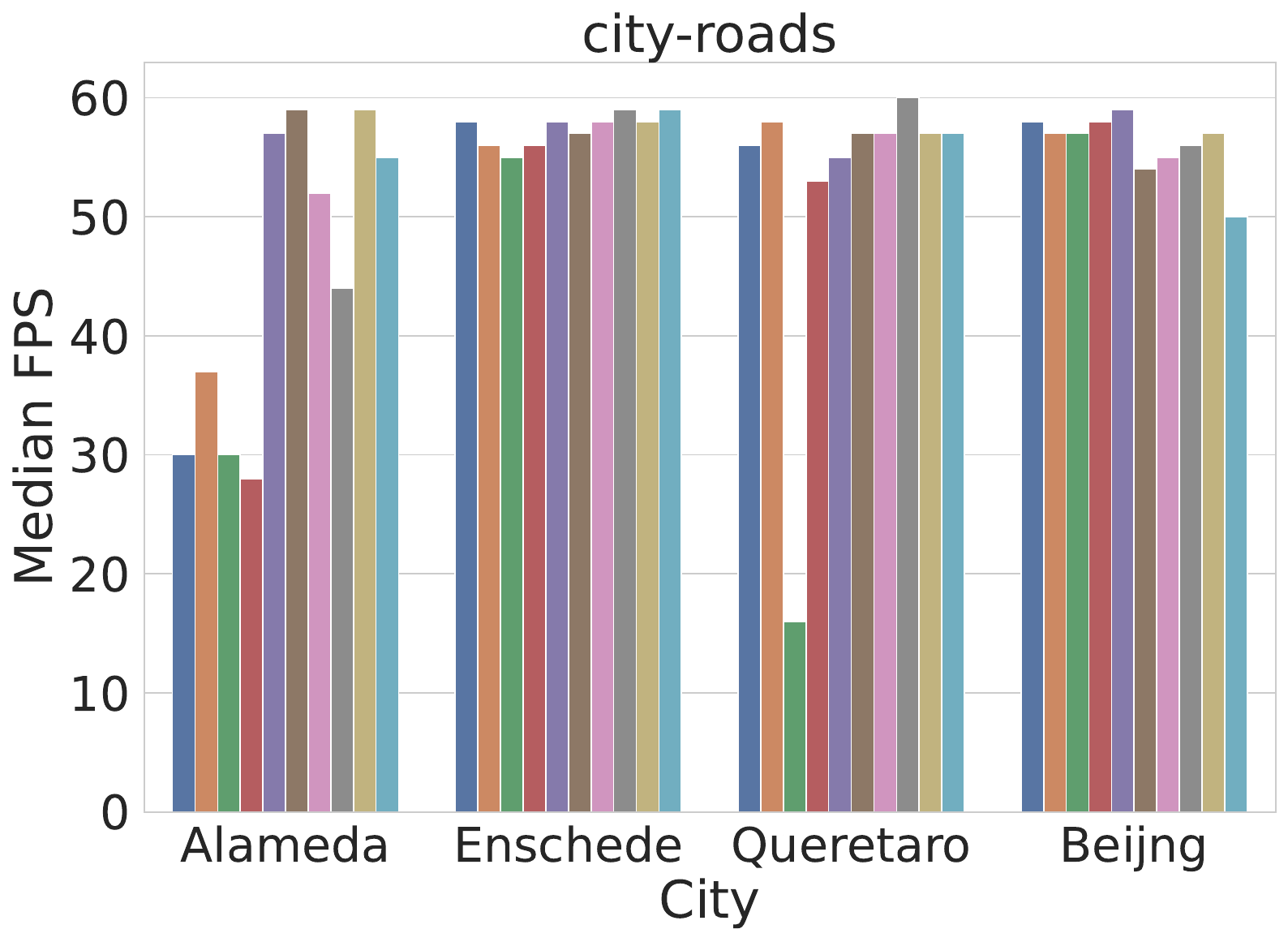}} 
 
\subfloat[]{\includegraphics[width=0.6\columnwidth]{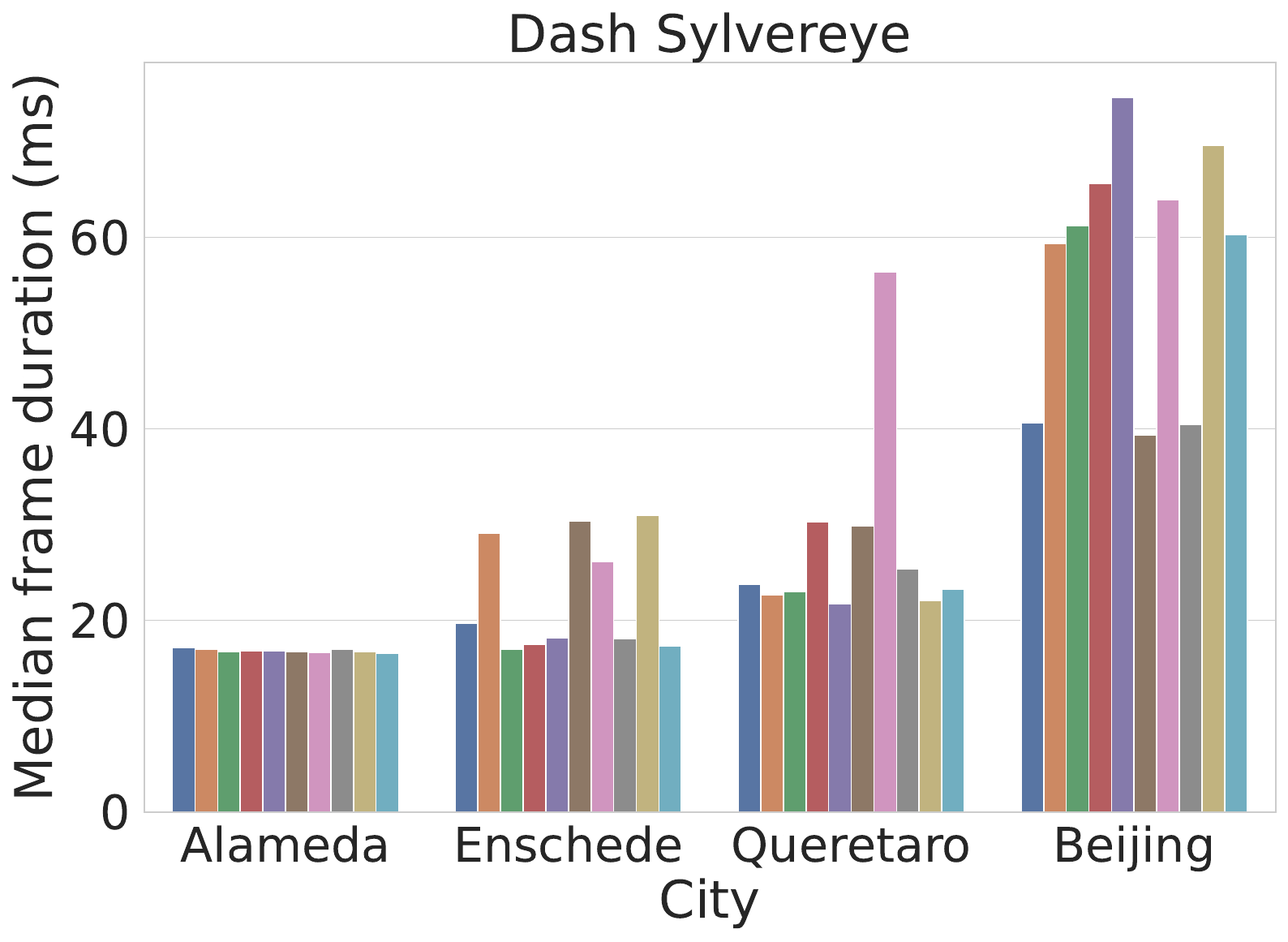}}\hspace{5mm}
\subfloat[]{\includegraphics[width=0.6\columnwidth]{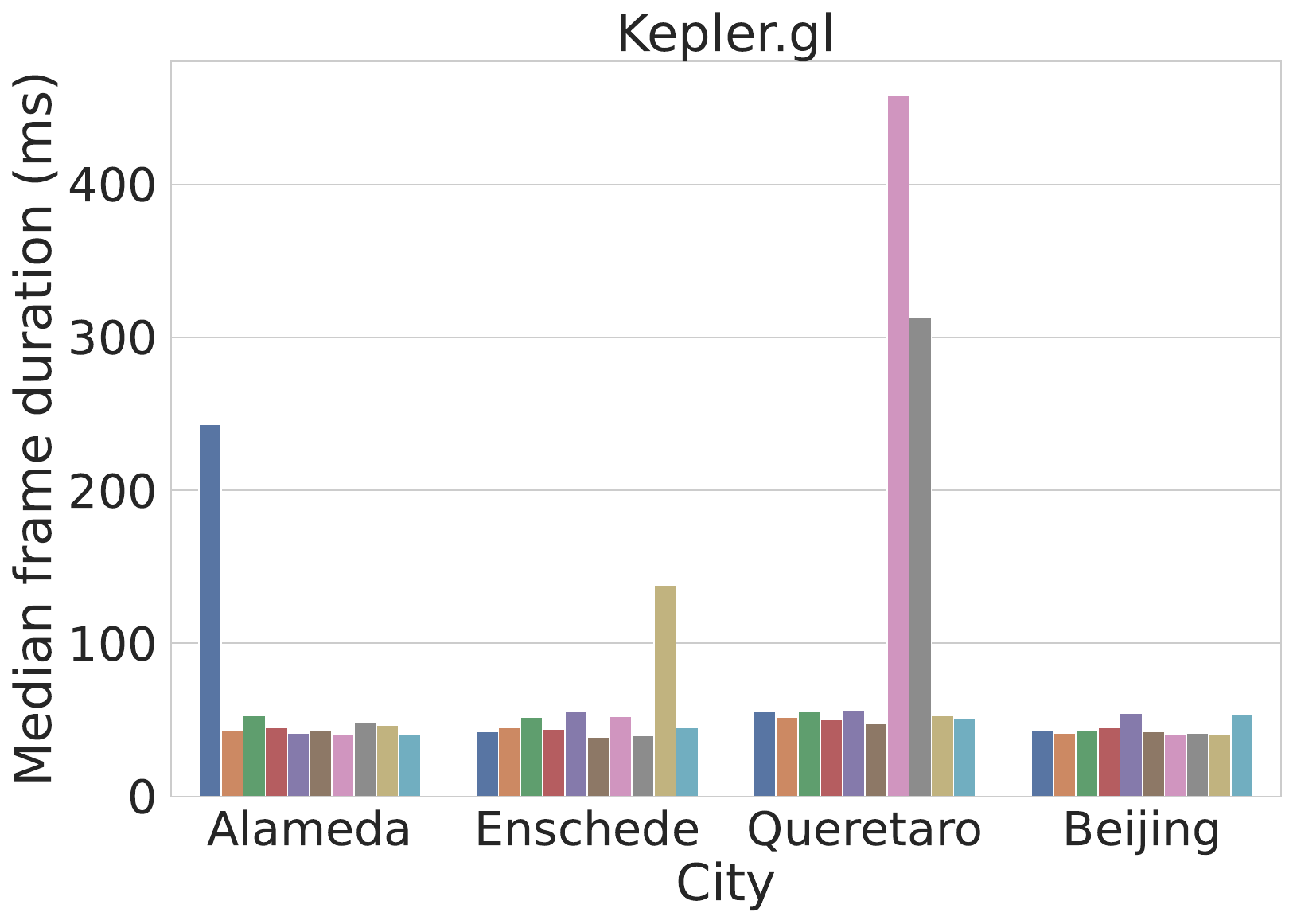}}\hspace{5mm}
\subfloat[]{\includegraphics[width=0.6\columnwidth]{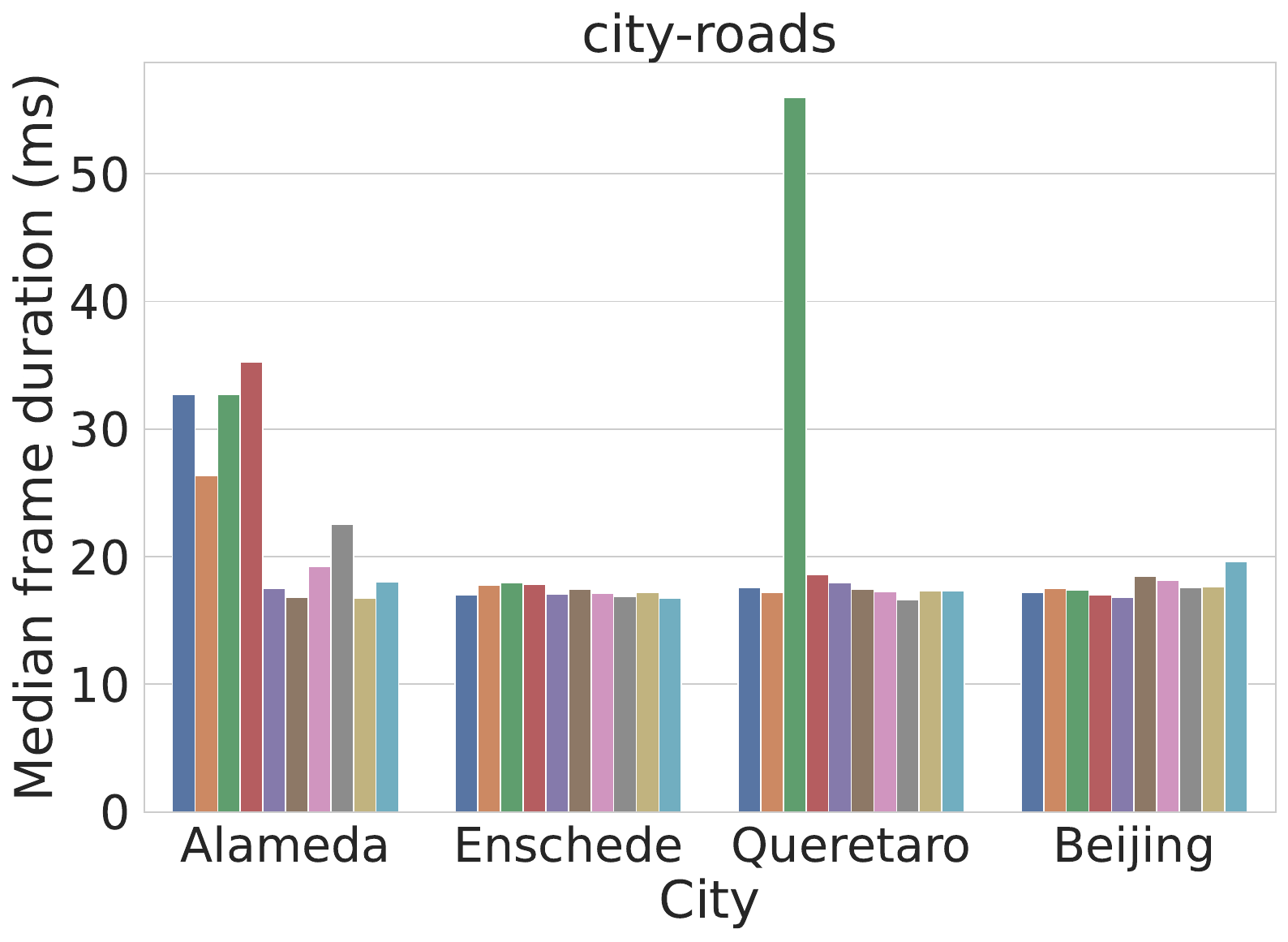}} 

\subfloat[]{\includegraphics[width=0.6\columnwidth]{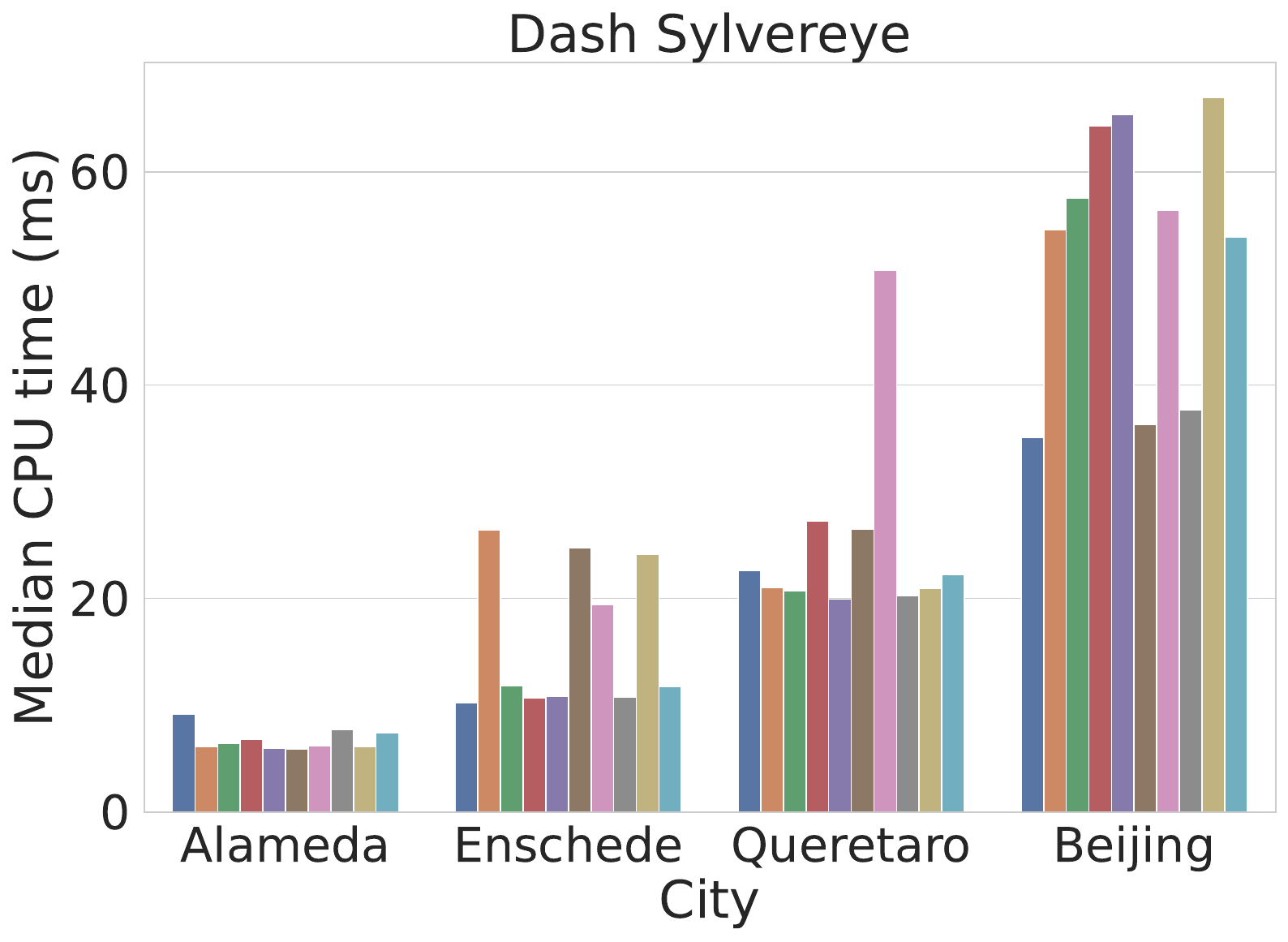}}\hspace{5mm}
\subfloat[]{\includegraphics[width=0.6\columnwidth]{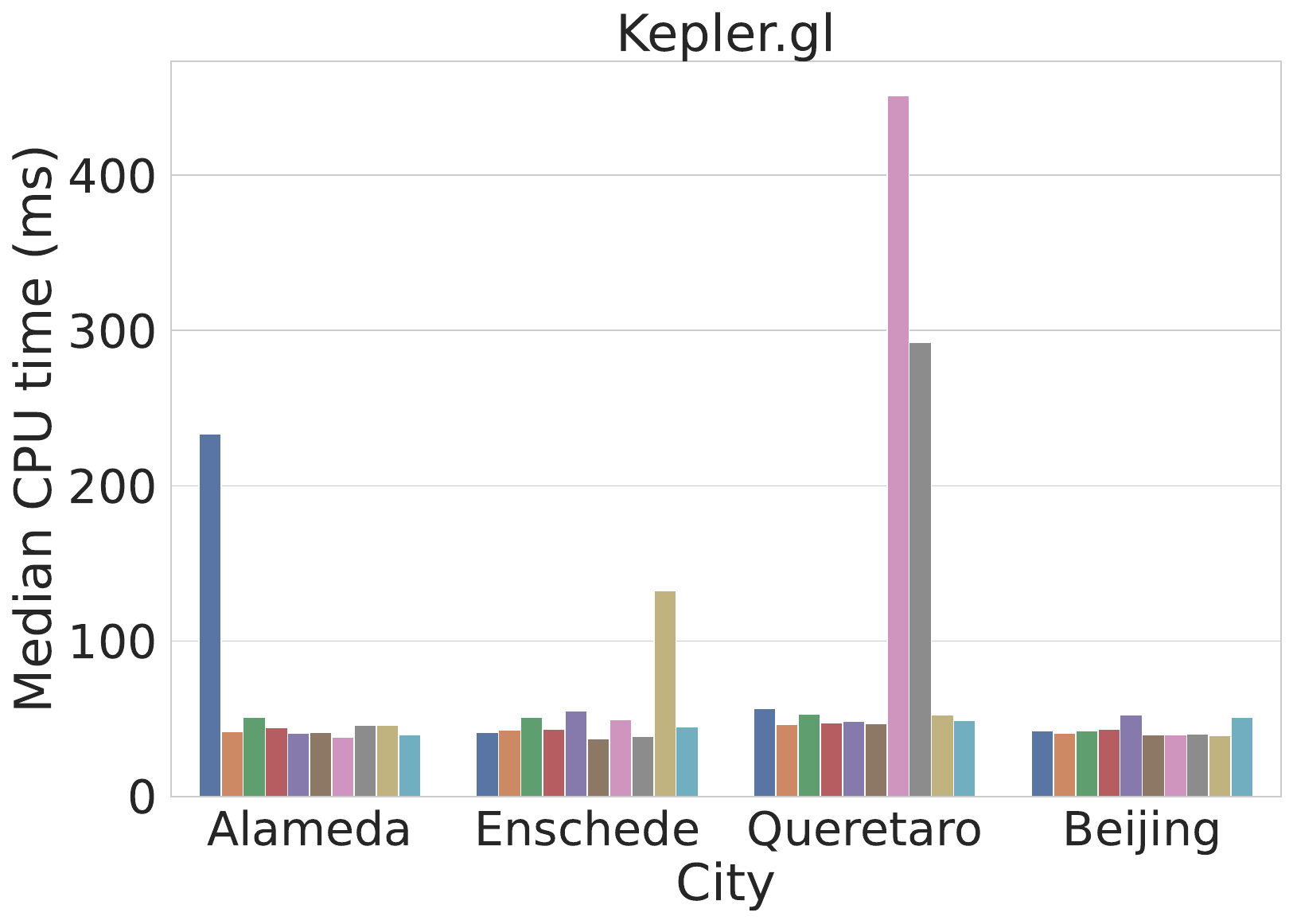}}\hspace{5mm}
\subfloat[]{\includegraphics[width=0.6\columnwidth]{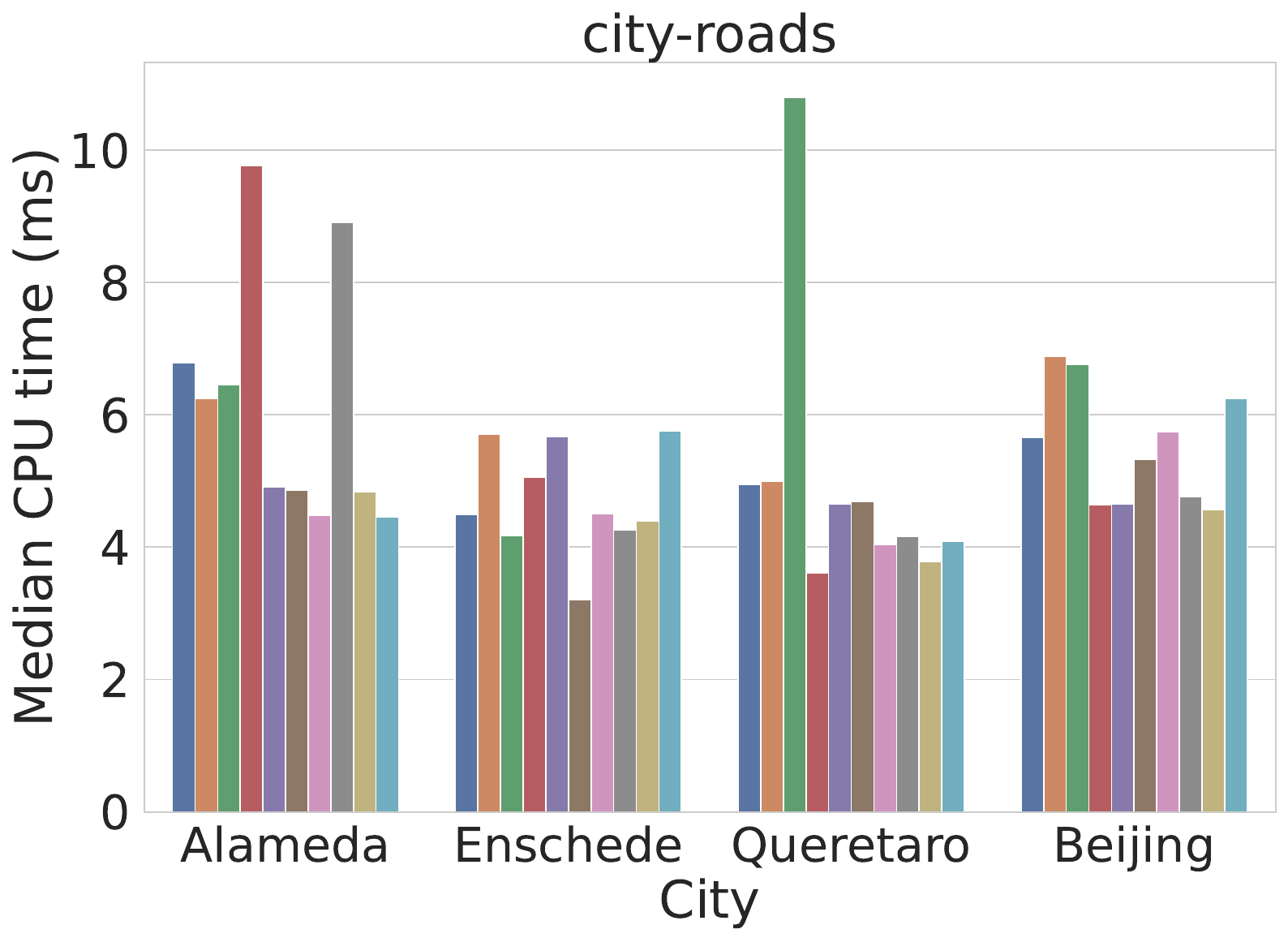}} 

\caption{Median frame FPS, duration, and CPU time shown by Dash Sylvereye, Kepler.gl, and city-roads on each experiment and each city. Cities are sorted from smaller to bigger from left to right.}
\label{fig:comparison-barplots}
\end{figure*}

\begin{figure*}  
\centering
\subfloat[]{\includegraphics[width=0.6\columnwidth]{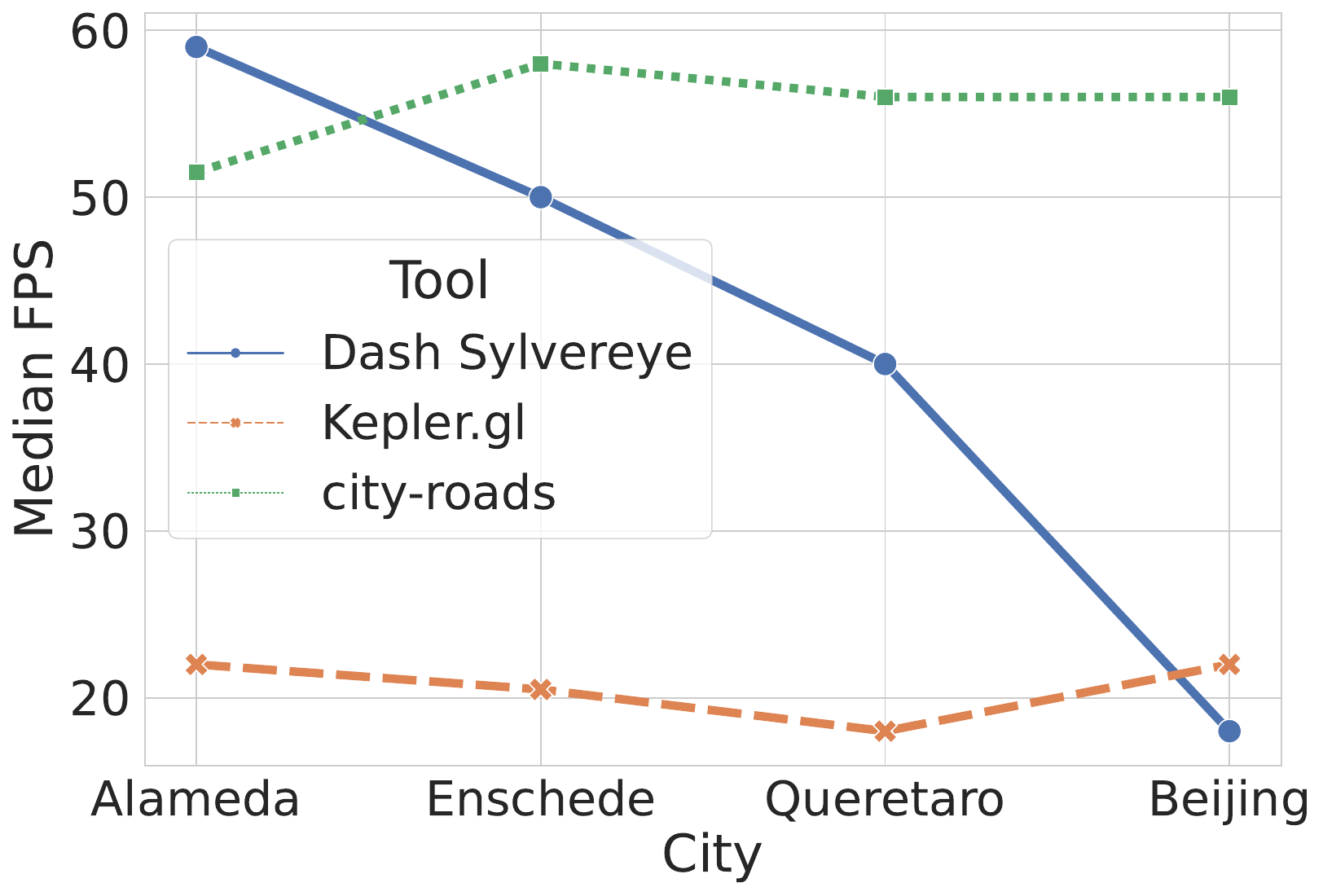}}\hspace{5mm}
\subfloat[]{\includegraphics[width=0.6\columnwidth]{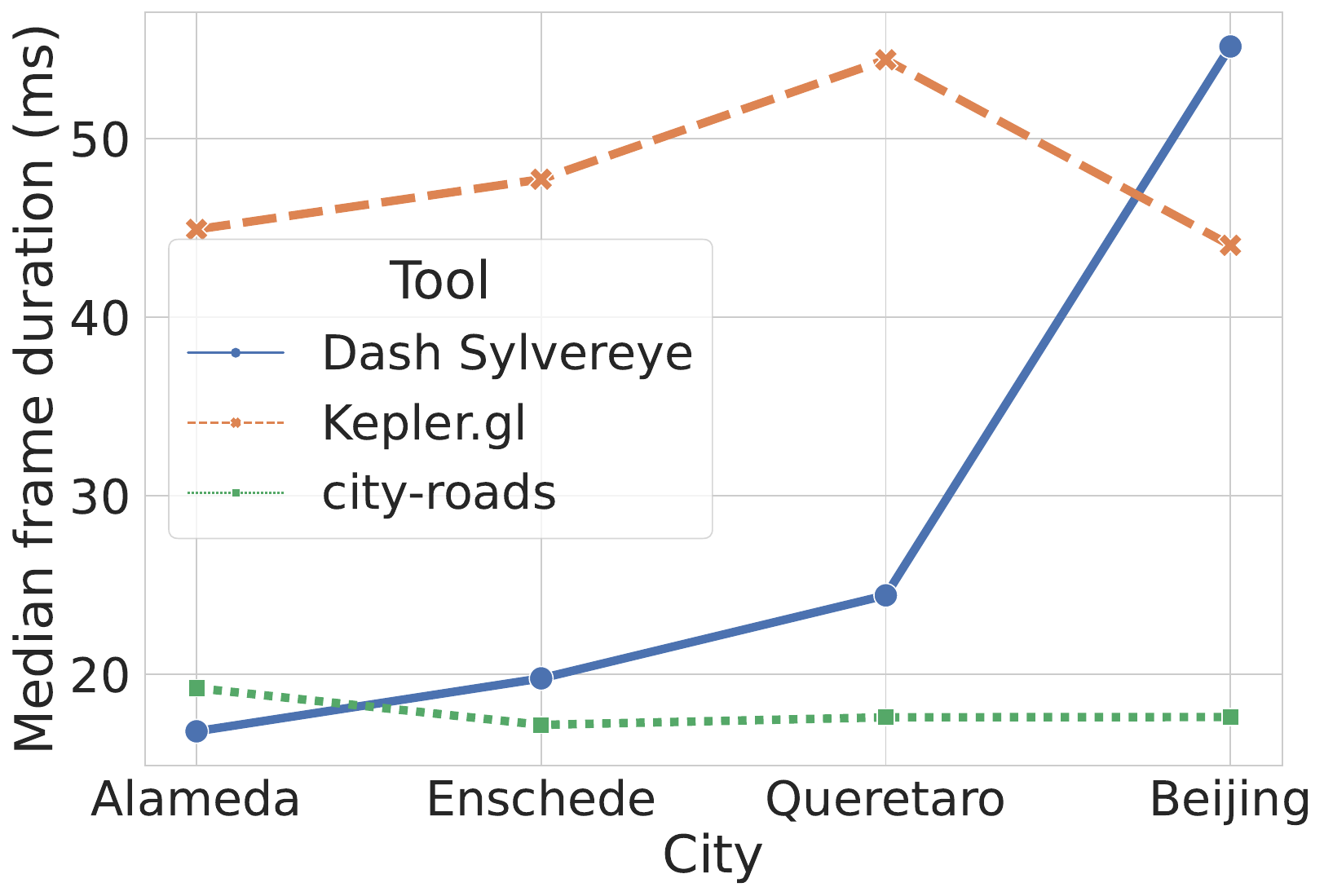}}\hspace{5mm}
\subfloat[]{\includegraphics[width=0.6\columnwidth]{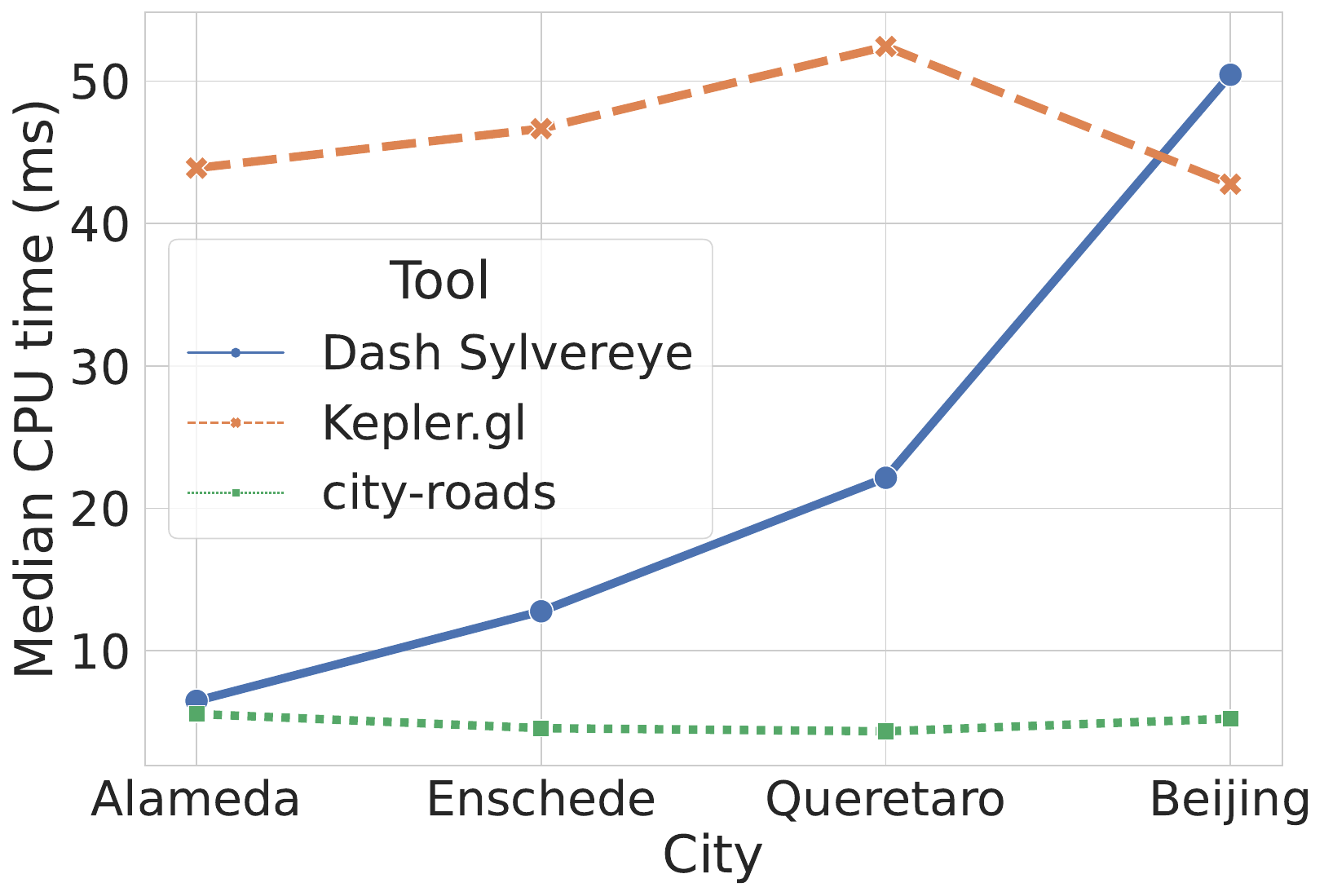}} 

\caption{Dash Sylvereye's median frame FPS, duration, and CPU time shown by Dash Sylvereye, Kepler.gl, and city-roads when merging all experiments for each city. Cities are sorted from smaller to bigger from left to right.}
\label{fig:comparison-lineplots}
\end{figure*}

Fig. \ref{fig:comparison-barplots} shows the median frame FPS, duration, and CPU time for each experiment and each city for the three tools. Fig. \ref{fig:comparison-lineplots} shows the median values when merging all experiments for each city for the three tools.

Figures show that, for the three tools, lower FPS values are associated with larger CPU times and frame durations. However, Figures also show that, unlike Dash Sylvereye, Kepler.gl and city-roads seem to be unaffected by the road network size. This suggests that these tools might exploit hardware acceleration during the panning process. In contrast, recall that Dash Sylvereye exploits hardware acceleration only after panning (and zooming) for redrawing.

Nonetheless, note in Fig. \ref{fig:comparison-barplots} that Kepler.gl showed FPSs lower than 10 in one experiment for Alameda, one experiment for Enschede, and two experiments for Queretaro. In contrast, for Alameda, Enschede, and Queretaro, Dash Sylvereye's FPS was higher than 20 whereas the duration and CPU remained low. For the largest city, Beijing, Dash Sylvereye's FPS was higher than 20 for three experiments.

Overall, Figures \ref{fig:comparison-barplots} and \ref{fig:comparison-lineplots} show that Dash Sylvereye's performance was inversely proportional to the road network size. However, Fig. \ref{fig:comparison-barplots} also shows that Dash Sylvereye FPS outperformed Kepler.gl in three out of four cities (Alameda, Enschede, and Queretaro), whereas it outperformed city-roads in one out of four cities (Alameda). Additionally, as city-roads, Dash Sylvereye showed FPSs above 24 for three out of four cities (Alameda, Enschede, and Queretaro). With these results, Dash Sylvereye showed to be competitive when compared to both Kepler.gl and city-roads for road networks with dozens of thousands of nodes and edges.
 
\ifheavyimages
\begin{figure*}
\centering
\includegraphics[scale=3]{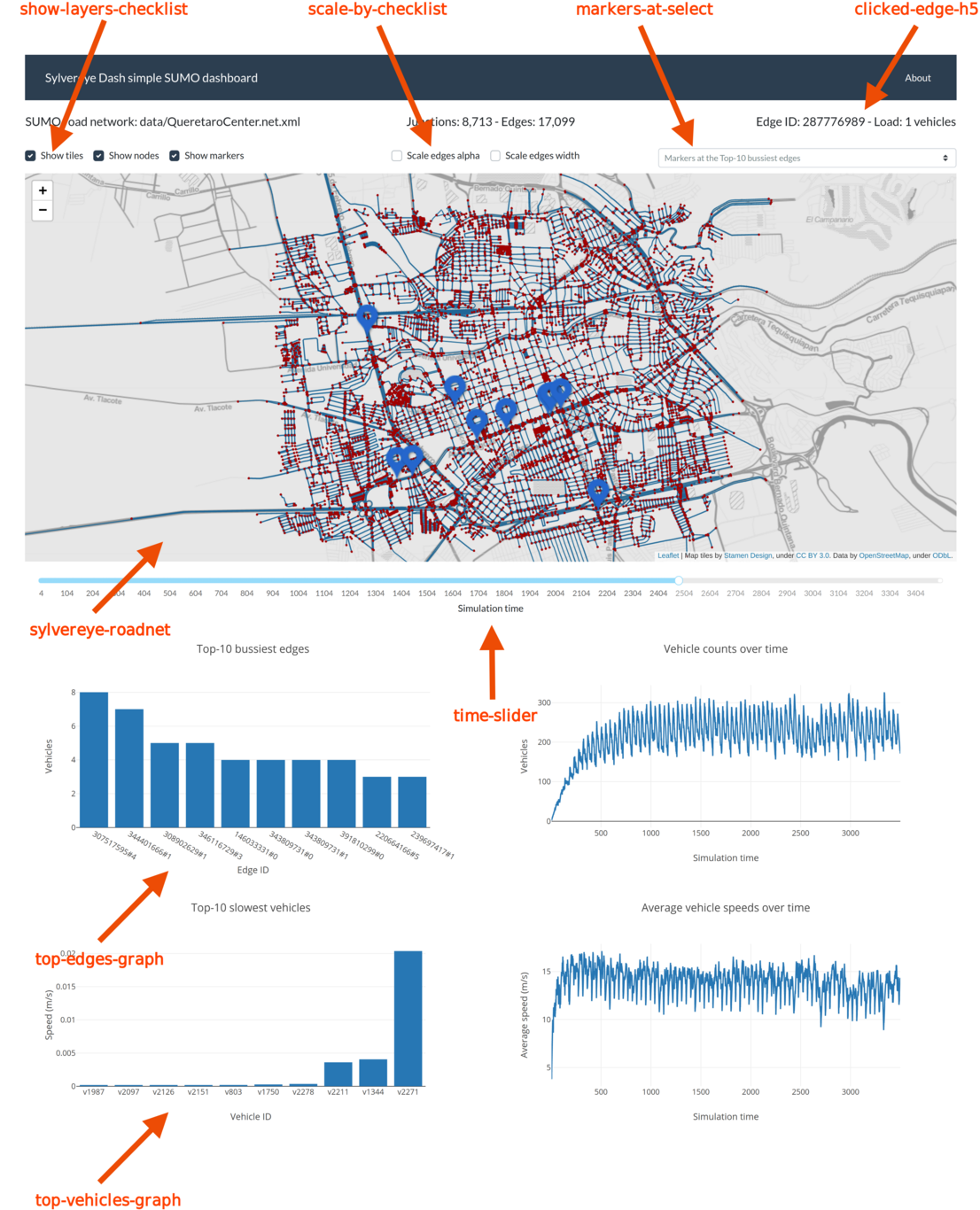} 

\caption{Layout of the dashboard made with Dash and the Dash Sylvereye library for analyzing a SUMO traffic simulation. Labels in orange are the Dash component identifiers referred to in Fig. \ref{fig:callbacks}.}
\label{fig:sumo-dashboard}
\end{figure*}
\fi


\ifheavyimages
\begin{figure*}
\centering
\includegraphics[scale=1.3]{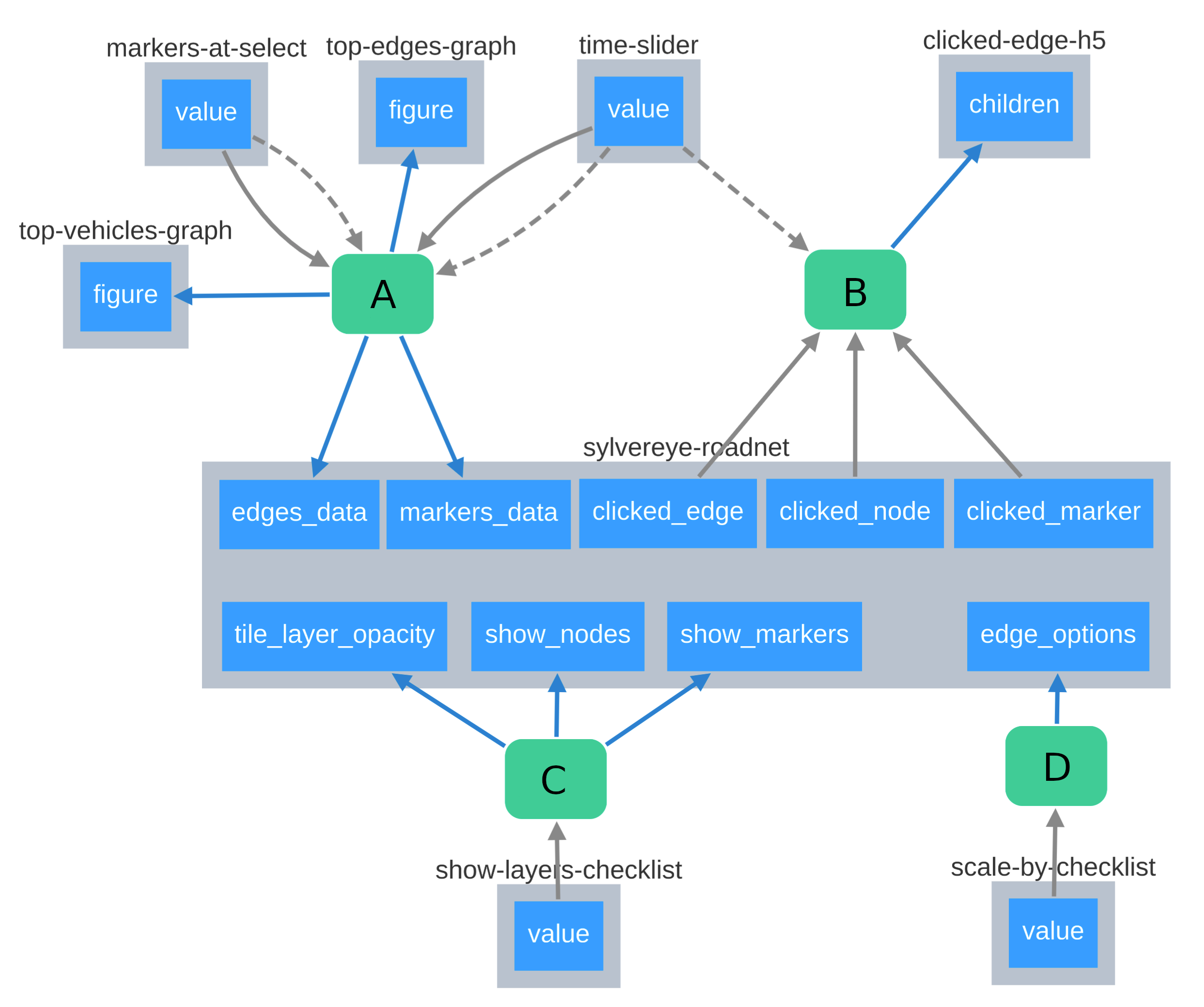}  

\caption{Callback graph of the SUMO simulator dashboard example, as generated by the Dash Dev Tools. Gray boxes represent Dash components. Labels on top of gray boxes are the Dash component identifiers. Green rounded boxes represent callback functions. Blue boxes represent the input and output properties. Solid gray arrows pointing to green rounded boxes come from input callback parameters. Solid blue arrows going out of green rounded boxes point to output callback parameters. Dashed gray arrows pointing to green circles represent states.}
\label{fig:callbacks}
\end{figure*}
\fi

\section{Dashboard example: Queretaro City traffic simulation}\label{section:dashboard-example}

This section presents the design and implementation of an example dashboard application written with the Dash framework that exploits Dash Sylvereye for the analysis of postmortem simulation data on the street network of Queretaro City, Mexico. For simulations, we made use of the SUMO urban traffic simulator \cite{Behrisch2011}, a well-known simulator in the field of urban analysis.

The purpose of this section is twofold. Firstly, we intend to better illustrate the usefulness of Dash Sylvereye in assisting a traffic analyst to observe how traffic bottlenecks build up with time in a busy transportation area made of thousands of street roads and junctions through a dashboard visualization centered around Dash Sylvereye. Secondly, we intend to offer the reader more details about how the Dash Slyvereye component can be integrated into a non-trivial dashboard by coordinating it with other charts and UI controls for multivariate data visualization.

\ifheavyimages
\begin{figure*}
\centering
\subfloat[]{\includegraphics[scale=2.2]{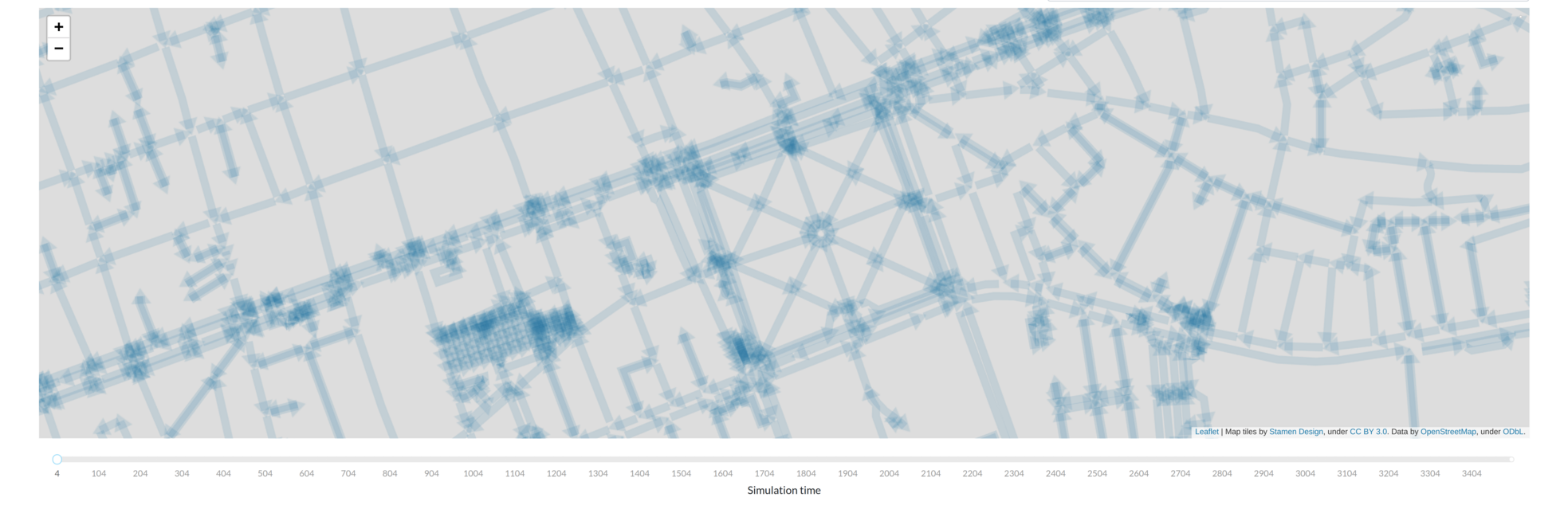}} 

\subfloat[]{\includegraphics[scale=2.2]{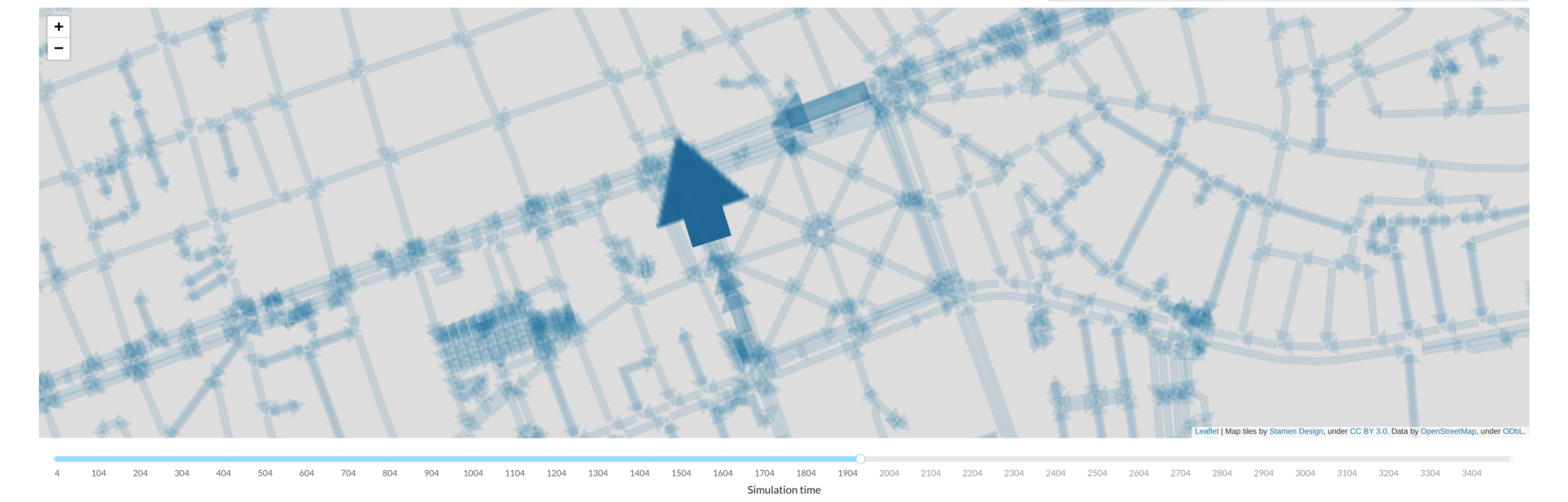}} 

\subfloat[]{\includegraphics[scale=2.2]{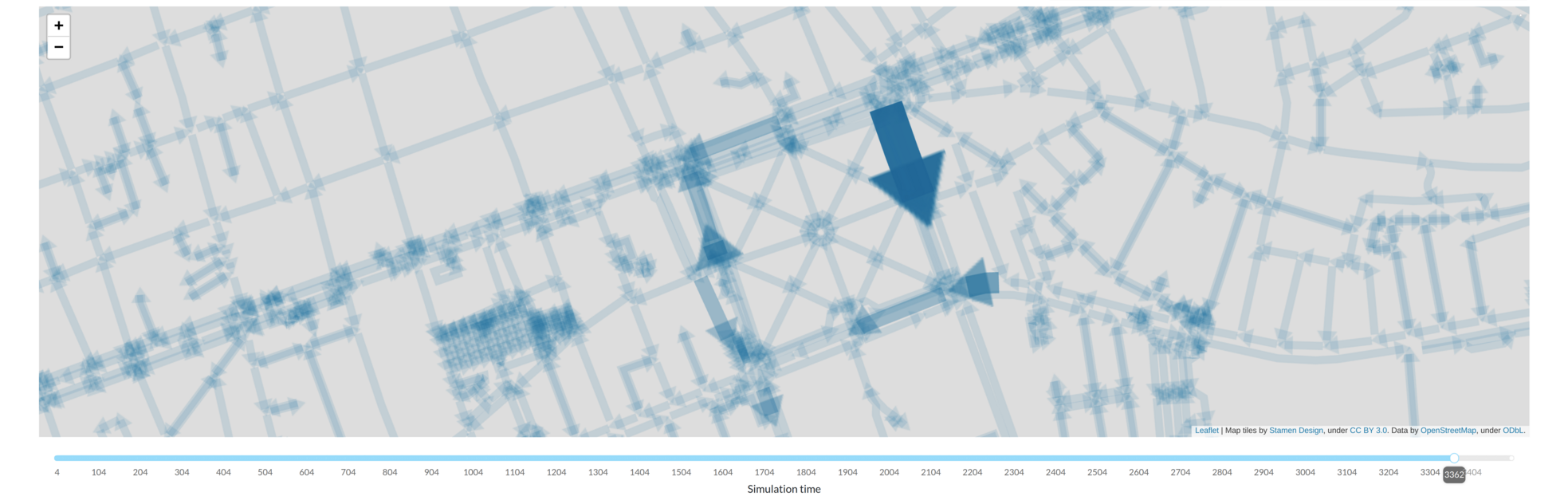}} 

\caption{Radiography-like visualizations centered at the ``Alameda Hidalgo'' park at three arbitrarily selected but consecutive simulation time steps. The visualization style was configured through the GUI components of the SUMO traffic dashboard developed with Dash and the Dash Sylvereye library. Note that the large blue arrows are street roads rendered by Dash Sylvereye (images were not edited).}
\label{fig:radiographies}
\end{figure*}
\fi

\subsection{Street network retrieval and simulation}

A street network from the center of the metropolitan area of Queretaro City, Mexico, was manually selected and downloaded in OSM format by taking advantage of the export features of the OSM website. The resulting street network had 8,713 nodes and 17,099 edges. The OSM network was converted to SUMO’s XML format by using SUMO’s netconvert tool. Finally, the SUMO XML network was converted to Dash Sylvereye’s list-of-dictionaries format with the help of the Sumolib Python library\footnote{https://sumo.dlr.de/docs/Tools/Sumolib.html}.

To create synthetic vehicle trip data, a simulation was run by using SUMO as follows: random trips for vehicles were generated with SUMO’s \verb|randomTrips.py| script. A SUMO simulation was run for 3,500 timesteps with the \verb|--fcd-output| flag to save the Floating Car Data (FCD) of all timesteps in XML format.  The produced FCD data was then processed to get CSV files that could be conveniently imported into the Dash dashboard application. CSV files included the vehicle count for edges at each timestep, the speed of vehicles at each timestep, total vehicle counts for each timestep, and average vehicle speed at each timestep. 

\subsection{Layout design}
 
Recall from Section \ref{section:background} that a Dash dashboard application is composed of two parts: 1) the layout which describes what the application looks like and  2) the callbacks that define the interactivity of the application. Fig. \ref{fig:sumo-dashboard} shows a screenshot of the resulting dashboard layout.

The dashboard includes a Dash Sylvereye visualization as its main element (\verb|sylvereye-roadnet|). The user can select which layers are visible through the \verb|show-layers-checklist| checklist. Markers are displayed at either the middle of edges with the highest vehicle counts or atop the slowest vehicles, depending on the option selected by the user in the \verb|markers-at-select| selection list. 

The user can also select which visual attributes of edges (transparency, width) to scale in proportion to the edge vehicle count through the \verb|scale-by-checklist| checklist. The dashboard shows a slider to allow the user to select the desired simulation time to display (\verb|time-slider|). When the user changes the simulation time, a callback is triggered to update:

\begin{itemize}
    \item The network edges, width, and transparency.
    \item The position and the popup texts of markers.
    \item A bar plot of the top-10 edges with the highest vehicle count in the network (\verb|top-edges-graph|)
    \item A bar plot of the speed for the top-10 slowest vehicles (\verb|top-vehicles-graph|)
\end{itemize}
 
When the user clicks either an edge, a node, or a marker, data about the clicked element is shown in the label at the top-right corner (\verb|clicked-edge-h5|). Finally, the dashboard also shows a line plot showing the vehicle count over time and a line plot showing the average vehicle speed over time. However, these two plots are static in the sense that they do not need to change as a result of the interaction of the user with the dashboard.

\subsection{Callback design}
 
Fig. \ref{fig:callbacks} shows the callback graph of the SUMO simulator dashboard example, as generated by the Dash Dev Tools. The application contains four main callbacks, callbacks A to D, which together define the interactivity of the whole application.

Callback A triggers when either: 
\begin{enumerate}
    \item The value of \verb|markers-at-select| changes its value, in which case the callback outputs a new set of markers.
    \item The \verb|time-slider| changes its value, in which case the callback updates the markers and both bar plots \verb|top-vehicles-graph| and \verb|top-edges-graph| to reflect the data at the new time step.
\end{enumerate}

Callback B triggers when any of the click attributes of the Dash Sylvereye component changes its value as the result of a user clicking on a node, an edge, or a marker. The callback updates the label \verb|clicked-edge-h5| with info about the clicked element.

Callback C triggers when \verb|show-layers-checklist| changes its value because the user selected a different set of layers to show. The callback updates the show/hide properties of the Dash Sylvereye component accordingly.

Finally, callback D triggers when the component \verb|scale-by-checklist| changes its value because the user selected different scale options. The callback updates the \verb|edge_options| attribute of the Dash Sylvereye component accordingly to update the edge's alpha and width style methods.

\subsection{Visualization insights}
 
Fig. \ref{fig:radiographies} shows screenshots of the Dash Sylvereye component at three arbitrarily selected simulation time steps. The tilemap is centered at the ``Alameda Hidalgo'' park, a centric place where traffic bottlenecks build up in real life. Edges transparency and width were scaled to the vehicle count by checking the corresponding checkboxes in the dashboard. The map tile layer, as well as the nodes and markers layers, were hidden by unchecking the corresponding checkboxes. The result was a radiography-like visualization of the vehicle traffic. The ``radiography'' in Fig. \ref{fig:radiographies} clearly shows that, even with random trips, vehicle traffic builds up on the main street roads surrounding the park.


\section{Conclusion}\label{section:conclusion}

This paper presented Dash Sylvereye, a new Python library for generating web-based visualizations of large street networks, delivered as a component for the widely-used Dash framework. To the best of our knowledge, Dash Sylvereye is the first tool written for Python that generates street network visualizations atop web tile maps that supports programmable user interactivity, that is designed as a component of a dashboard framework from the ground up, and that supports WebGL. Dash Sylvereye can be combined with other Dash UI and chart components to enable the development of interactive dashboard visualizations around street network data.

We showed that Dash Sylvereye can offer fast response speeds (close to 60 FPS) for street networks with thousands of edges. We also found Dash Sylvereye to be competitive when compared to the state-of-the-art visualization libraries Kepler.gl and city-roads for road networks with dozens of thousands of nodes and edges. With the help of a dashboard application example, we explored how Dash Sylvereye can be utilized as a convenient tool for interactively analyzing multivariate traffic data.

Visualization generation time is an important factor that impacts the experience of the end-user. Even with WebGL acceleration, we have observed that the visualization first drawing and redrawing of very large graphs in Dash Sylvereye may take non-negligible time on a commodity system, an overhead not present in other libraries like Kepler.gl and city-roads. This overhead includes the time needed for the generation of the graphics (sprites and polygons) of the street network and the computation of hit polygons for edge click detection. Future work includes methodologically assessing visualization generation times on commodity computers and evaluating optimization options.

Similar to other web-based visualization tools, one of Dash Sylvereye’s main drawbacks is that the size of a street network the library can handle is limited by the system’s physical memory and the GPU memory capacity. In this regard, an interesting research venue is to study efficient graph coarsening algorithms for edge bundling that 1) allow the tool to handle very large networks and 2) help the researcher’s cognitive process of making sense of such complex structures. 

We plan to release Dash Sylvereye under an open-source license, enabling anyone to use it for their specific street network visualization needs.


\begin{IEEEbiography}[{\includegraphics[width=1in,height=1.25in,clip,keepaspectratio]{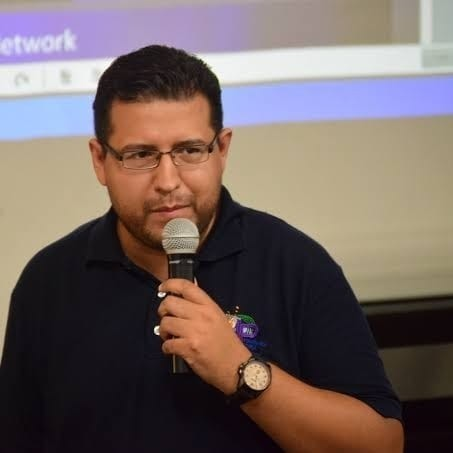}}]{Alberto Garcia-Robledo} holds a MSc. and a Ph.D. in Computer Science from the Center for Research and Advanced Studies of the National Polytechnic Institute (Mexico). He was a tech lead at the Geospatial Data Center of the Massachusetts Institute of Technology (US). Currently, he is a Conacyt Research Fellow at the Center for Research in Geography and Geomatics (Mexico). His current research interests include HPC, Big Data, Graph Analytics, and Visual Analytics. 
\end{IEEEbiography}

\begin{IEEEbiography}[{\includegraphics[width=1in,height=1.25in,clip,keepaspectratio]{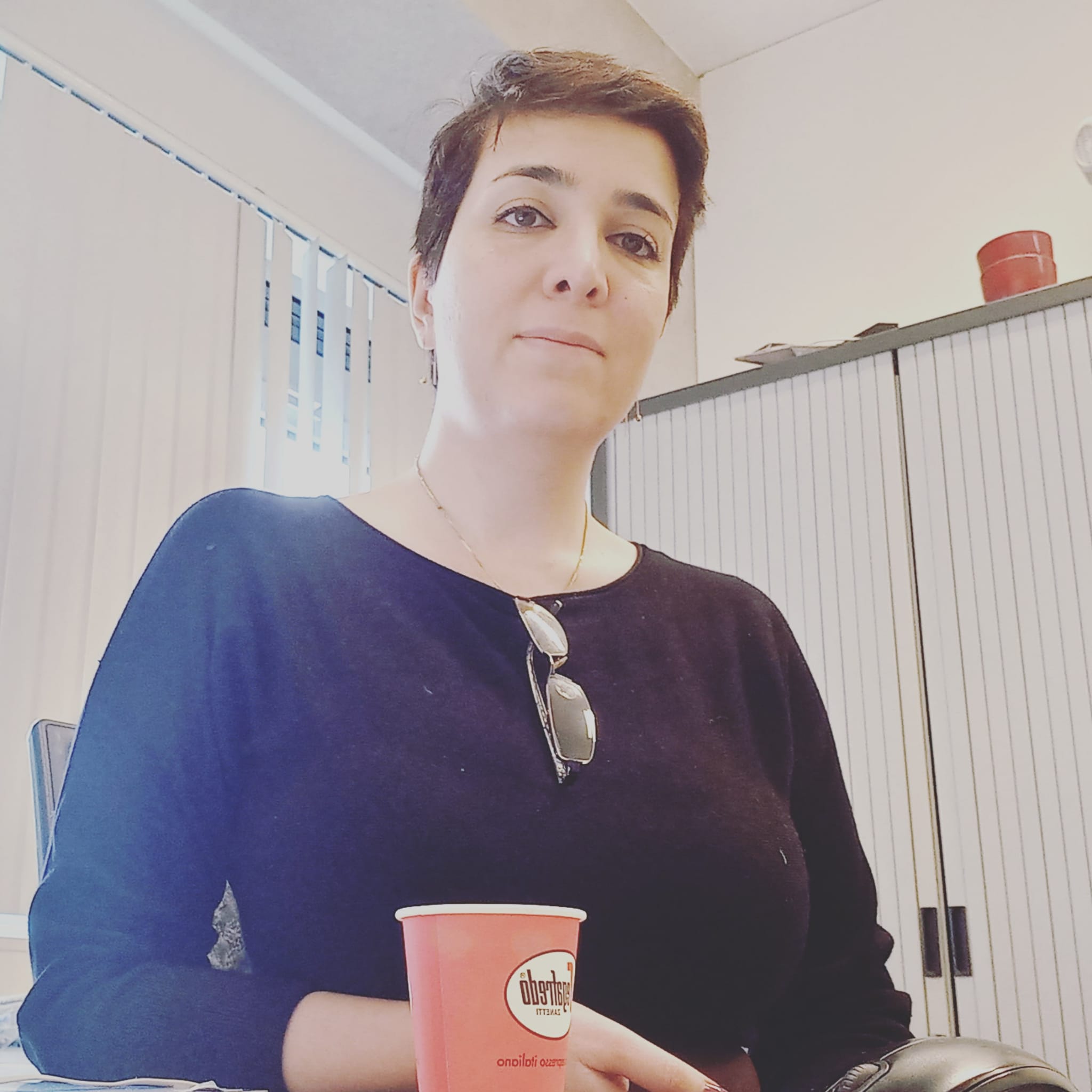}}]{Mahboobeh Zangiabady} is a lecturer at the Design and Analysis of Communication Systems (DACS) group at the University of Twente. She holds a Ph.D. in computer science from the Centre for Research and Advanced Studies of the National Polytechnic Institute. Her research interests cover network virtualization, QoS, resource management, machine learning, and Network Functions Virtualization, Software-Defined Networks (NFV/SDN).
\end{IEEEbiography}

\EOD

\end{document}